\documentclass{aa}  
\usepackage{graphicx}
\usepackage{amsmath}
\usepackage{amsfonts}
\usepackage{amssymb}
\usepackage{gensymb}
\usepackage{textcomp}
\usepackage[varg]{txfonts}
\usepackage[T1]{fontenc}
\usepackage{natbib}
\usepackage{epstopdf}
\usepackage{ulem}
\bibpunct{(}{)}{;}{a}{}{,}

\newcommand{\hri}{HRI$_{\textrm{EUV}}$}

\begin{document}

\title{Solar coronal heating from small-scale magnetic braids}

\author{L.~P.~Chitta\inst{1}, 
H.~Peter\inst{1}, 
S.~Parenti\inst{2}, 
D.~Berghmans\inst{3}, 
F.~Auch\`{e}re\inst{2}, 
S.~K.~Solanki\inst{1,4},
R.~Aznar Cuadrado\inst{1}, 
U.~Sch\"{u}hle\inst{1},  
L.~Teriaca\inst{1}, 
S.~Mandal\inst{1}, 
 K.~Barczynski\inst{5,6}, 
\'{E}.~Buchlin\inst{2}, 
L.~Harra\inst{5,6}, 
E.~Kraaikamp\inst{3},
D.~M.~Long\inst{7},
L.~Rodriguez\inst{3},
C.~Schwanitz\inst{5,6}, 
P.~J.~Smith\inst{7},
C.~Verbeeck\inst{3}, 
A.~N.~Zhukov\inst{3,8},
W.~Liu\inst{9,10},
\and
M.~C.~M.~Cheung\inst{9}
}
\institute{Max-Planck-Institut f\"ur Sonnensystemforschung, Justus-von-Liebig-Weg 3, 37077 G\"ottingen, Germany\\
\email{chitta@mps.mpg.de}
\and
Universit\'{e} Paris-Saclay, CNRS, Institut d'astrophysique spatiale, 91405, Orsay, France
\and
Solar-Terrestrial Centre of Excellence - SIDC, Royal Observatory of Belgium, Ringlaan -3- Av. Circulaire, 1180 Brussels, Belgium
\and
School of Space Research, Kyung Hee University, Yongin, Gyeonggi 446-701, Republic of Korea
\and
Physikalisch-Meteorologisches Observatorium Davos, World Radiation Center, 7260 Davos Dorf, Switzerland 
\and
ETH-Z\"{u}rich, Wolfgang-Pauli-Str. 27, 8093 Z\"{u}rich, Switzerland
\and
UCL-Mullard Space Science Laboratory, Holmbury St. Mary, Dorking, Surrey RH5 6NT, UK
\and
Skobeltsyn Institute of Nuclear Physics, Moscow State University, 119992 Moscow, Russia
\and
Lockheed Martin Solar and Astrophysics Laboratory, 3251 Hanover Street, Building 252, Palo Alto, CA 94304, USA
\and
Bay Area Environmental Research Institute, NASA Research Park, Moffett Field, CA 94035, USA
}

   \date{Received ; accepted }

\abstract
{Relaxation of braided coronal magnetic fields through reconnection is thought to be a source of energy to heat plasma in active region coronal loops. However, observations of active region coronal heating associated with an untangling of magnetic braids remain sparse. One reason for this paucity could be the lack of coronal observations with a sufficiently high spatial and temporal resolution to capture this process in action. Using new observations with high spatial resolution (250--270\,km on the Sun) and high cadence (3--10\,s) from the Extreme Ultraviolet Imager (EUI) on board Solar Orbiter, we observed the untangling of small-scale coronal braids in different active regions. The untangling is associated with impulsive heating of the gas in these braided loops. We assess that coronal magnetic braids overlying cooler chromospheric filamentary structures are perhaps more common. Furthermore, our observations show signatures of spatially coherent and intermittent coronal heating during the relaxation of the magnetic braids. Our study reveals the operation of gentle and impulsive modes of magnetic reconnection in the solar corona.}

   \keywords{Sun: corona --- Sun: magnetic fields --- Magnetic reconnection --- Plasmas}
   \titlerunning{Solar coronal heating from small-scale magnetic braids}
   \authorrunning{L. P. Chitta et al.}

   \maketitle

\section{Introduction\label{sec:int}}
One of the widely invoked models to explain the million Kelvin solar corona is magnetic braiding \citep[][]{1972ApJ...174..499P,1983ApJ...264..642P}. In this scenario, magnetic footpoints of coronal loops are slowly stressed and randomly moved by photospheric convective motions, leading to a braiding of the magnetic field in the corona, that is, a wrapping of the field lines around each other. Current sheets form at these braids due to a misalignment of the magnetic field. These current sheets dissipate impulsively through reconnection, on timescales shorter than typical coronal cooling timescales, to produce storms of nanoflares throughout an entire loop, which then intermittently heat the coronal plasma to the observed temperatures in excess of 1\,MK \citep[][]{1988ApJ...330..474P,2006SoPh..234...41K}. This nanoflare heating from an untangling of the braids is postulated to be central for explaining active region loops that reach temperatures above 2\,MK. Photospheric convective motions also excite magnetohydrodynamic (MHD) waves, but observations and models show that these waves are not energetic enough to power these hot active region coronal loops \citep[][]{2011Natur.475..477M,2020SSRv..216..140V}. However, their contribution cannot be ruled out \citep[e.g.,][]{2019ApJ...881..107A} because at the photospheric level, the waves carry enough Poynting flux to heat the active region corona. The open issue here is whether this energy is channeled into the corona, and how waves might efficiently dissipate the energy to heat coronal plasma \citep[][]{2020SSRv..216..140V}.

Three-dimensional (3D) MHD models of the solar atmosphere do produce 1--4\,MK hot active region loops in the corona \citep[e.g.,][]{2002ApJ...572L.113G,2019A&A...624L..12W}. Here, the heating is generally ascribed to the slow stressing of photospheric footpoints. High-resolution 3D MHD simulations that include self-consistent convective motions induce both a slow stressing of the footpoints and waves into the corona \citep[][]{2022A&A...658A..45B}. However, the partition of the energy flux that is channeled into the corona by these two modes in this high-resolution model has not yet been fully explored.

Overall, the observational validation of coronal heating through nanoflares as proposed by \citet[][]{1988ApJ...330..474P} remains elusive. In particular, spatial, temporal, and thermal regimes in which magnetic braiding and the associated nanoflare heating operate in the solar corona are poorly constrained observationally. 

One of the early, but occasional indications of magnetic braiding or wrapping of loops in the corona was made through TRACE 171\,\AA\ filter images in the extreme ultraviolet (EUV) \citep[][]{1999SoPh..187..261S}. However, \citet[][]{2011ApJ...736....3V} argued that the observed TRACE cases of braiding were ambiguous \citep[see also][]{2017ApJ...837..108P}, and that most TRACE loops fail to show evidence of magnetic braiding. Furthermore, \citet[][]{2011ApJ...736....3V} examined X-ray images of solar active region loops and found that loops appeared to cross each other in only a few sites. From this, they concluded that braiding, if present on the Sun, should exist on smaller transverse scales, corresponding to the angular (spatial) scales of less than 5\arcsec. Using EUV and X-ray observations of an active region, \citet{2010A&A...517A..41P} observed the relaxation and multithermal evolution of tangled loops in a postflare arcade. The lack of overwhelming evidence for magnetic braiding in the X-ray observations might be due to their generally lower spatial resolutions, which are lower than 1\arcsec. Despite resolving 1\arcsec\ on the Sun, TRACE observations did not reveal widespread coronal braiding. Coronal braiding might be observable on angular scales smaller than 1\arcsec.

\begin{table*}
\setlength{\tabcolsep}{4pt}
\begin{center}
\caption{Details of observations recorded by the EUI/\hri\ instrument on board Solar Orbiter.\label{tab:table}}
\begin{tabular}{c c c c c c c c c c}
\hline\hline
Target & Date \& start time & Distance  & Scale & Cadence & Duration & \multicolumn{2}{c}{Carrington longitude\tablefootmark{e}} & NOAA & Overview\\
\cline{7-8}
ID &  (UT)\tablefootmark{a} & (AU)\tablefootmark{b} &  (km)\tablefootmark{c} & (s) & (min)\tablefootmark{d} & Orbiter (\textdegree) & Earth (\textdegree) & AR number\tablefootmark{f} & Figure\\
\hline
I   & 2022-03-17 03:23:08 & 0.379 & 135 & 3 & 45 & 259 & 233 & 12965 & Fig.\,\ref{fig:over1}\\
II  & 2022-03-19 10:41:20 & 0.356 & 127 & 5 & 60 & 238 & 202 & 12967  & Fig.\,\ref{fig:over2}\\
III \& IV & 2022-04-01 09:24:41 & 0.347 & 124 & 10 & 75 & 136 & 32 & 12975 \& 12976 & Fig.\,\ref{fig:over3} \\
\hline
\end{tabular}

\tablefoot{
\tablefoottext{a}{Start time of observations corrected for the time delay between Earth and Solar Orbiter.}
\tablefoottext{b}{Distance of Solar Orbiter from the Sun in astronomical units.}
\tablefoottext{c}{Image scale of \hri\ in km per pixel. The spatial resolution is about twice this value.}
\tablefoottext{d}{Duration of the observational sequence in minutes.}
\tablefoottext{e}{Carrington longitude of Solar Orbiter and Earth.}
\tablefoottext{f}{NOAA active region number of the observed target.}
}
\end{center}
\end{table*}

\begin{figure}
\begin{center}
\includegraphics[width=0.49\textwidth]{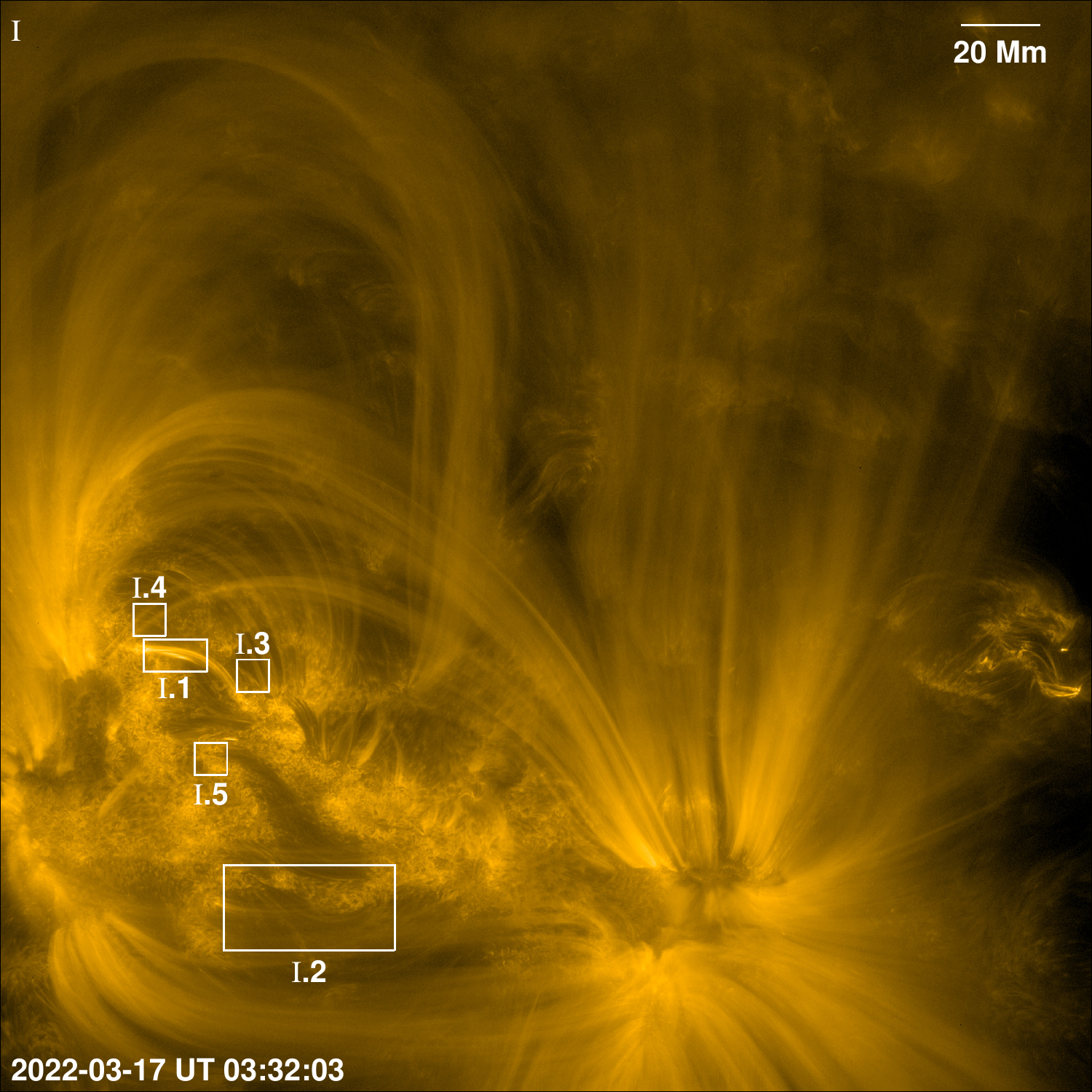}
\caption{Overview of AR12965 (AR-I) as observed with EUI/\hri\ on 2022 March 17. The snapshot shows the full field of view of the \hri\ covering AR-I. See Table\,\ref{tab:table} for observational details. Boxes I.1--I.5 cover different regions exhibiting intermittent heating events in the corona that we analyzed. Close-up views of regions I.1 and I.2 are displayed in Figs.\,\ref{fig:braid1} and \ref{fig:braid2}, respectively. Regions I.3--I.5 are displayed in Fig.\,\ref{fig:bright}. The intensity of the displayed image is on a log scale. North is approximately up. See Sects.\,\ref{sec:obs} and \ref{sec:braid} and Appendix\,\ref{sec:app} for a discussion.
\label{fig:over1}}
\end{center}
\end{figure}

To this end, using high-resolution (0.4\arcsec -- 0.5\arcsec), high-cadence ($\sim$5\,s) coronal EUV observations recorded by the High-resolution Coronal Imager sounding rocket experiment \citep[Hi-C;][]{2014SoPh..289.4393K}, \citet{2013Natur.493..501C} found signatures of spatially resolved magnetic braids and the associated coronal heating in two cases of low-lying loops in an active region. They implied that the observable signatures of coronal braiding would be common at small spatial scales of about 150\,km. Magnetic field extrapolations revealed that one of the braided loops in Hi-C images is associated with a low-lying twisted flux rope \citep[][]{2014ApJ...780..102T} above a penumbral filament region. These loops are not associated with the main opposite-polarity patches of this active region. The other example of small braided loops in Fig.\,3 of \citet{2013Natur.493..501C} might be due to so-called moss regions in this active region, as suggested by \citet{2014ApJ...787...87V}. Because Hi-C was flown on a suborbital rocket flight, the duration of observations it was able to record was limited to only some 5\,minutes. Despite this limitation, Hi-C set a benchmark for high-resolution and high-cadence coronal observations by revealing the highly dynamic subarcsecond structure of the solar corona.

More recently, \citet{2021NatAs...5...54A} observed signatures of nanojets and accompanying coronal heating in loops undergoing episodes of coronal rain. The authors suggested that nanojets are produced as a result of magnetic reconnection and subsequent relaxation of braided coronal loops. Furthermore, \citet{2022ApJ...934..190S}, who observed additional cases of nanojets in magnetic features hosting cooler plasma including coronal rain, suggested that nanojets are a general result of magnetic reconnection and that not only braiding, but also Kelvin–Helmholtz and Rayleigh–Taylor instabilities can cause them. An open question in nanojets is whether they occur in coronal loops that do not show a signature of cool plasma or coronal rain. Therefore, more examples of braided coronal fields, especially in normal coronal loops, are urgently required. 

We present evidence of magnetic braiding and the associated intermittent coronal heating in four active regions using novel high-resolution, high-cadence observations by the 174\,\AA\ EUV High Resolution Imager (\hri) of the Extreme Ultraviolet Imager \citep[EUI;][]{2020A&A...642A...8R} on board the Solar Orbiter mission \citep[][]{2020A&A...642A...1M}. The data were obtained during the first perihelion observing campaigns of EUI (see Table\,\ref{tab:table} for details). With its full width at half maximum (FWHM) of point spread function (PSF) equivalent to $\sim$0.35\arcsec\ as seen from Earth, EUI/\hri\ resolves the structures in the solar corona during this period as well as or better than Hi-C \citep[][]{2014SoPh..289.4393K,2019SoPh..294..174R}. In addition, some of the EUI observations provide not only a higher cadence, but also a longer time coverage.

\section{Observations\label{sec:obs}}

The thermal response function of the \hri\ instrument has its peak at temperatures of about 1\,MK due to Fe\,{\sc ix} (at 171.1\,\AA) and Fe\,{\sc x} (at 174.5\,\AA\ and 177.2\,\AA). In this study, we employ \hri\ data from three recent first perihelion observing campaigns covering active regions. Relevant information of the \hri\ observing campaigns are summarized in Table\,\ref{tab:table}. The PSF FWHM of \hri\ is about twice the image scale per pixel. For the observed period, the spatial resolution of \hri\ is thus in the range of 250--270\,km on the Sun. We used the calibrated level-2 \hri\ data from these campaigns \citep[][]{euidatarelease5}. We removed the spacecraft jitter in the level-2 \hri\ data using a cross-correlation technique (see Appendix\,\ref{sec:align} for details).

For the observing campaign on 2022 March 17 covering AR12965 (AR-I), we complemented EUI/\hri\ data with the cotemporal EUV observations from the Atmospheric Imaging Assembly \citep[AIA;][]{2012SoPh..275...17L} on board the Solar Dynamics Observatory \citep[SDO;][]{2012SoPh..275....3P}. In particular, we used the data from the six EUV passbands of SDO/AIA. The central wavelengths of these AIA filters, dominant ion species contributing to the emission in active regions and log$_{10}T$\,(K) formation temperature of the ions are 94\,\AA\ (Fe\,{\sc x}: 6.05; Fe\,{\sc xviii}: 6.85), 131\,\AA\ (Fe\,{\sc viii}: 5.6; Fe\,{\sc xxi}: 7.05),  171\,\AA\ (Fe\,{\sc ix}: 5.85), 193\,\AA\ (Fe\,{\sc xii}: 6.2; Fe\,{\sc xxiv}: 7.25), 211\,\AA\ (Fe\,{\sc xiv}: 6.3), and 335\,\AA\ (Fe\,{\sc xvi}: 6.45) \citep[see][]{2010A&A...521A..21O,2012SoPh..275...41B}. We processed the level-1 AIA data using the standard \texttt{aia$\_$respike} and \texttt{aia$\_$prep} procedures available in the solarsoft library \citep[][]{1998SoPh..182..497F}. The AIA data have an image scale of 0.6\arcsec\,pixel$^{-1}$. During this period, SDO/AIA was operating in a special mode to support the EUI/\hri\ campaign, in which its 131\,\AA, 171\,\AA,\ and 193\,\AA\ passbands recorded data at a cadence of 6\,s, while the 94\,\AA, 211\,\AA,\ and 335\,\AA\ filters recorded data every 96\,s. The snapshot of the AIA 171\,\AA\ image from this period in Fig.\,\ref{fig:aia} shows the view of AR-I from the vantage point of Earth.

\begin{figure*}
\begin{center}
\includegraphics[width=\textwidth]{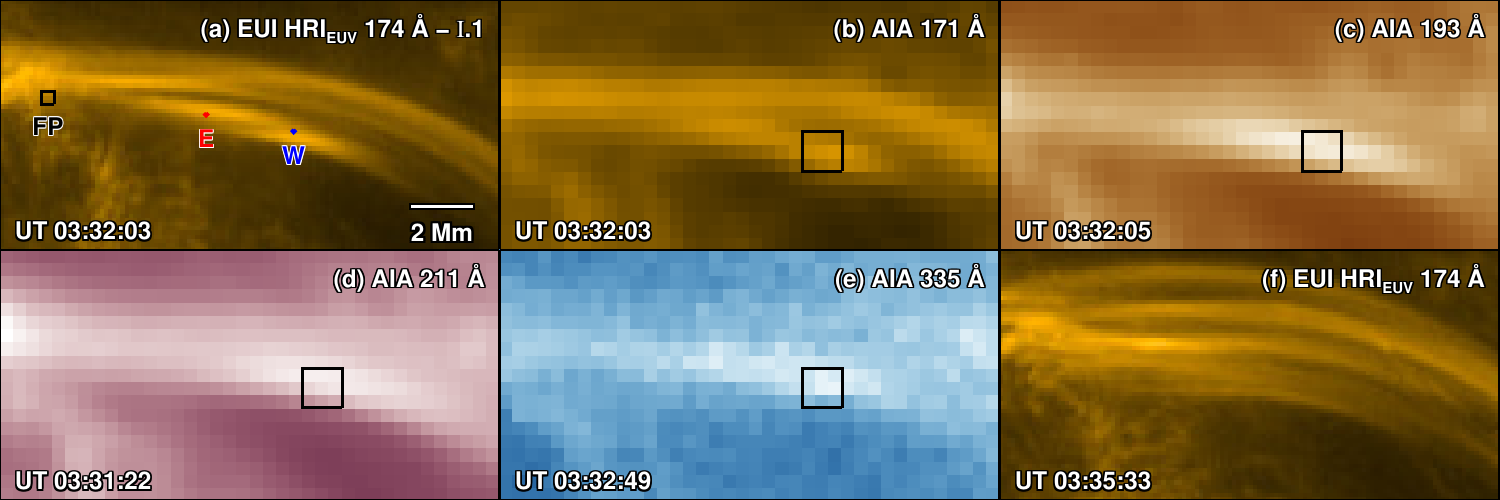}
\caption{Braided coronal loops. Panel (a) displays a segment of braided coronal loops observed in the core of AR-I with \hri\ (box I.1 in Fig.\,\ref{fig:over1}). Red and blue pixels, labeled E and W, mark the eastern and western sections of the braided loops, while the black box FP is placed at the footpoint region of these loops. The time-stamp of the snapshot and the spatial scale (using a 2\,Mm bar) are also indicated. Panels (b)--(e) display the same braided loops from the perspective of SDO/AIA in different EUV passbands as labeled. The projected area of these panels is roughly the same as that of panel (a). The black box in panels (b)--(e) is overlaid on the location of the braided loops. The mean emission of the respective AIA passbands computed from the region covered by this black box is plotted in Fig.\,\ref{fig:lc}. In panel (f), the \hri\ snapshot shows the relaxed state of the initially braided loops, only about 3 minutes after panel (a); see also Fig.\,\ref{fig:seq} for the detailed evolution of the braided loops. The intensities of all the snapshots are on a square-root-scale. See Sect.\,\ref{sec:braid} for a discussion. An animation of panel a is available online.
\label{fig:braid1}}
\end{center}
\end{figure*}

This active region was also observed by the X-Ray Telescope \citep[XRT;][]{2007SoPh..243...63G} on board the Hinode mission \citep[][]{2007SoPh..243....3K}. We used the data from the Al$-$poly and Be$-$thin filters of Hinode/XRT. These images were first deconvolved with the model PSF of XRT and were then processed with the \texttt{xrt$\_$prep} procedure that is distributed in solarsoft. The XRT data have an image scale of about 1\arcsec\,pixel$^{-1}$ and a cadence of 30\,s. The details of extracting features of interest in AR-I from \hri, AIA, and XRT observations are provided in Appendix\,\ref{sec:coal}.

For the remaining two EUI campaigns (i.e., on 2022 March 19 and 2022 April 01; see Table\,\ref{tab:table}), we did not make use of complementary coronal SDO/AIA and or Hinode/XRT data mainly for simplicity. Nevertheless, to understand the connection between the coronal braids and the chromosphere, we did consider AIA 304\,\AA\ data that reveal cooler chromospheric and transition region structures around log$_{10}T$\,(K) of 4.7 sampled by He\,{\sc ii}. These AIA 304\,\AA\ data were also processed in the same way as above (but we skipped the respike step for simplicity).

Overall, the four active regions considered in this study are representative of active regions that exhibit weak to moderate flaring activity (see Appendix\,\ref{sec:app} for a related discussion). The time-stamps of \hri\ observations quoted are corrected for the time delay between Earth and Solar Orbiter\footnote{Hinode is in a Sun-synchronous orbit around Earth at an altitude of about 650\,km. The SDO is in an inclined geosynchronous orbit around Earth at an altitude of about 35.756$\times$10$^{3}$\,km. The light travel time from these telescopes to Earth is in the range of 0.002\,s to 0.12\,s. This is more than three orders of magnitude shorter than the light travel time between Solar Orbiter and Earth. For simplicity, we therefore consider the Earth time itself as reference to correct the timestamps of \hri\ images.}.

\section{Braided loops in the core of an active region\label{sec:braid}}

Using \hri\ observations, we detected a localized brightening in a system of coronal loops\footnote{We refer to a well-defined arch-like EUV feature as a loop. Many such loops then form a system of loops.} in the core of AR-I on 2022 March 17 around UT\,03:32 (see box AR-I.1 in Fig.\,\ref{fig:over1}). The loops themselves appear about 13\,minutes after a small B-class flare in AR-I (see Fig.\,\ref{fig:goes}), and thus they could be considered as post-flare loops. Upon a closer examination, we found that the localized brightening is associated with a pair of coronal loops that were entwined or braided. The braiding is most clearly seen between the locations marked E and W in Fig.\,\ref{fig:braid1}a. Although the braiding is not clear in SDO/AIA data because of its limited spatial resolution, different AIA passbands do show local enhancement in the EUV emission at the location where EUI sees the braiding (Fig.\,\ref{fig:braid1}b--e). The braided loops relax over the course of the next 3\,minutes and give rise to coronal loops that run more parallel to each other. Simultaneously, the localized brightening also fades away (Figs.\,\ref{fig:braid1}f and \ref{fig:seq}). We note that only this pair of the loops that show apparent braiding exhibit localized brightening. All the other loops observed with \hri\ that formed in the aftermath of this B-class flare show no apparent braiding signatures and exhibit rather uniform brightening along their length, unlike the event that we discussed. The localized brightening is also consistent with the  small-angle reconnection scenario of braided loops and nanojets \citep[][]{2021NatAs...5...54A}.

It might be argued that this apparent braided feature is actually a line-of-sight superposition of emission from two noninteracting loops. For this to be the case, however, there should be two loops with intersected localized brightenings in the line of sight that fade in a similar way. This ad hoc arrangement of noninteracting loops that exhibit coherent evolution is, however, unlikely. It might also be argued that EUV absorption due to the raining of cooler material could give the impression of braiding. However, we do not find any signatures of hotter loops in the direction of the initial dark feature (between points E and W in Fig.\,\ref{fig:braid1}a). At the same time, there is no evidence of any raining blobs either at that location in the AIA 304\,\AA\ passband (sensitive to cooler plasma). This supposed EUV absorption feature again requires a special arrangement of loops that is not supported by observations. Therefore, we suggest that the observed feature is indeed a braided system of coronal loops.

\begin{figure}
\begin{center}
\includegraphics[width=0.49\textwidth]{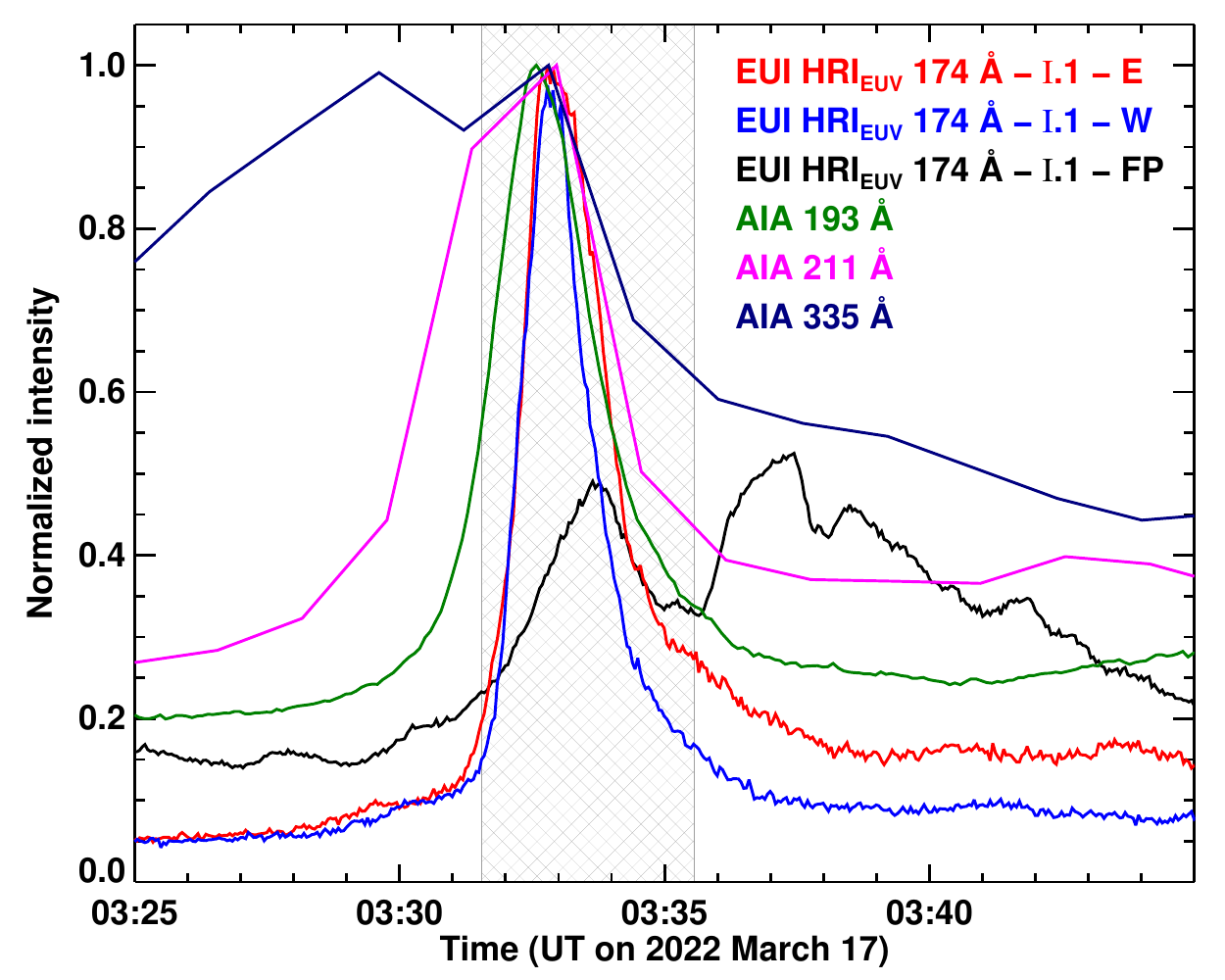}
\caption{Extreme ultraviolet emission characteristics of the braided loops. We plot the normalized EUV intensity from the \hri\ and SDO/AIA instruments. The red and blue light curves show the time evolution of EUV emission from braided loops at the pixels marked E and W in Fig.\,\ref{fig:braid1}a. The black curve is the average intensity from the footpoint region of the braided loops, labeled FP, in Fig.\,\ref{fig:braid1}a. The SDO/AIA light curves (dark green: 193\,\AA; magenta: 211\,\AA; dark blue: 335\,\AA) show the mean EUV intensity computed from the black boxes overlaid on the braided loops in the respective passband panels in Fig.\,\ref{fig:braid1}. The hatched region spans the duration covered by Fig.\,\ref{fig:seq}, which shows the relaxation of braided loops. See Sect.\,\ref{sec:braid} for a discussion.
\label{fig:lc}}
\end{center}
\end{figure}

The localized brightening at the site of braiding appears coherent over length scales of 2--5\,Mm. To investigate the evolution of coronal emission from distinct regions of the braided loops, we considered two pixels, marked E and W in Fig.\,\ref{fig:braid1}a, one on each of the two braided loops in the localized brightening region. These pixels are separated by more than 2\,Mm; that is, about seven times the spatial resolution of \hri\ at that time. The \hri\ intensities from these two pixels show simultaneous increases for roughly 120\,s. Thereafter, the intensity decreases monotonically, although more rapidly around W than around E (red and blue light curves in Fig.\,\ref{fig:lc}). The appearance of braided loops together with the localized intensity increase followed by the relaxation of the loops and the decrease in intensity is consistent with the scenario that magnetic energy stored in the coronal braids is released to heat the plasma on timescales of about 3\,minutes.

The AIA EUV diagnostics of this region shed further light on the thermal evolution associated with relaxation of coronal braids, even though the internal structure cannot be resolved. As we identified in Fig.\,\ref{fig:braid1}, the AIA EUV filters detected the localized brightening associated with the braided loops. We considered a region overlaid on this brightening and derived the mean intensity as a function of time in the six AIA EUV passbands, excluding the AIA 304\,\AA\ filter (black squares in Fig.\,\ref{fig:braid1}b--e). The light curves from three of the six EUV filters are plotted in Fig.\,\ref{fig:lc}. The thermal responses of the 193\,\AA, 211\,\AA,\ and 335\,\AA\ filters peak at progressively higher temperatures, that is, at log$_{10}T$\,(K) of 6.2, 6.3, and 6.45, respectively. We found that the light curves from these filters exhibit a progressively earlier rise phase with increasing temperature, and all of them start rising in intensity before \hri, which has a thermal response peak around log$_{10}T$\,(K) of 6. This suggests that the localized brightening associated with heating at the site of coronal braiding is caused by multithermal plasma. We further quantified this multithermal nature of plasma using a differential emission measure analysis (discussed in Appendix\,\ref{sec:dem}). This sequential signal of emission in filters sensitive to progressively cooler temperatures is thought to be a signature of plasma cooling in coronal loops \citep[][]{2012ApJ...753...35V}. However, we observed that the intensity evolution is very rapid on timescales of $\sim$200\,s compared to the longer cooling timescales, on the order of 1000\,s, as reported by \citet[][]{2012ApJ...753...35V}.

Furthermore, we visually identified the footpoint closer to the braiding feature. This footpoint region shows a localized intensity enhancement around 03:34\,UT, following the peak in \hri\ intensity (black curve in Fig.\,\ref{fig:lc}). In the vicinity of the visually identified footpoint, we also observed similar localized intensity enhancements, but shifted in time, that are likely associated with the other overlying loops. This type of footpoint intensity enhancement might be related to the deposition of energy in the transition region from a heating event. When the footpoint is associated with the braided loop system, the heating event might be linked to the energy released from the relaxation of the observed braids.

To conclude, we see clear evidence for the existence of braids within a system of coronal loops and for their subsequent relaxation. The timing of the emission originating at different temperatures indicates a rapid heating and multithermal evolution of the plasma in the loop in response to the relaxation of the braids.

\begin{figure*}
\begin{center}
\includegraphics[width=\textwidth]{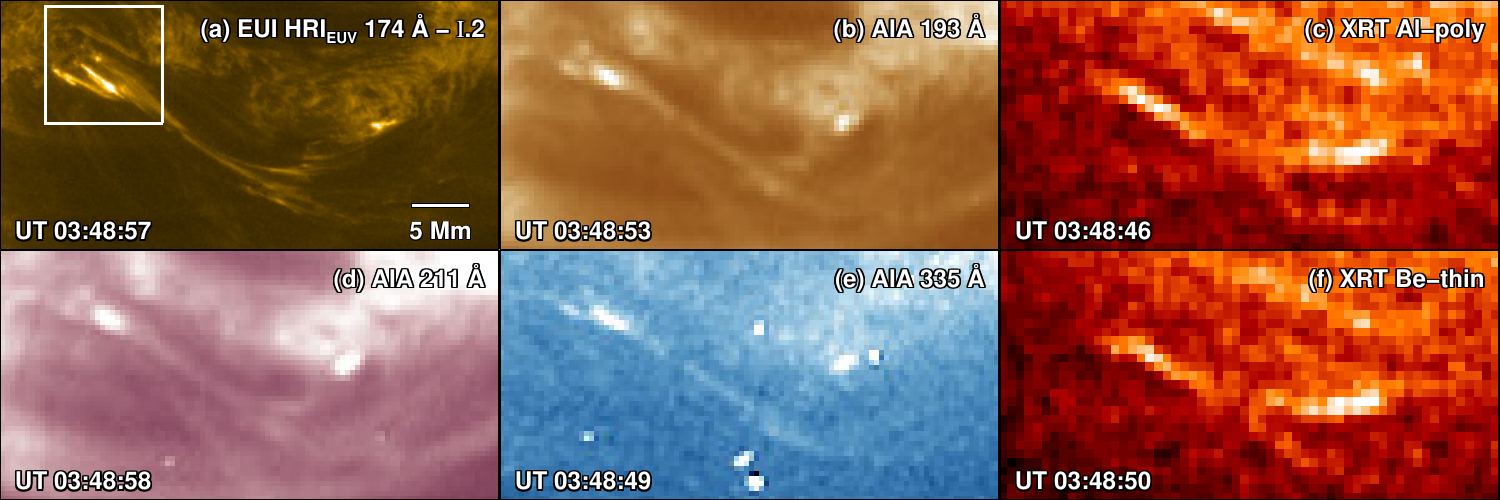}
\includegraphics[width=\textwidth]{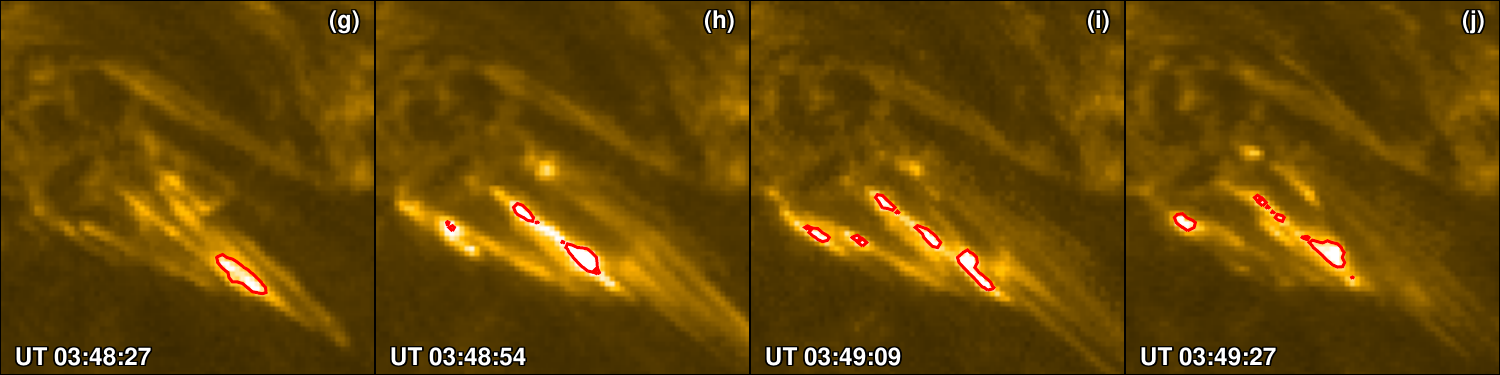}
\caption{Untangling of braided loops and coronal heating over  filamentary structures. Panel (a) is an \hri\ snapshot that captures coronal loops with complex braided structuring over cooler filaments in the southern portion of the core of AR-I (see box AR-I.2 in Fig.\,\ref{fig:over1}). The white box outlines the footpoint regions of the loops. Panels (b), (d), and (e) are snapshots covering roughly the same region as in panel (a), but  with EUV passbands of SDO/AIA as indicated. Panels (c) and (f) are nearly simultaneous snapshots from a pair of soft X-ray channels of Hinode/XRT. Panels (g)--(j) show a zoomed-in view of the footpoint regions of the loops as seen with \hri\ (white box in panel a). The red contours outline apparent blob-like features that are observed near the footpoints of the braided loops. The contours enclose regions that exceed a certain intensity threshold in each snapshot and are overlaid mainly to guide the eye. All the snapshots are displayed on a square-root scale. See Sect.\,\ref{sec:rapid} for a discussion. Animations of panel a and panel g are available online.
\label{fig:braid2}}
\end{center}
\end{figure*}

\begin{figure}
\begin{center}
\includegraphics[width=0.49\textwidth]{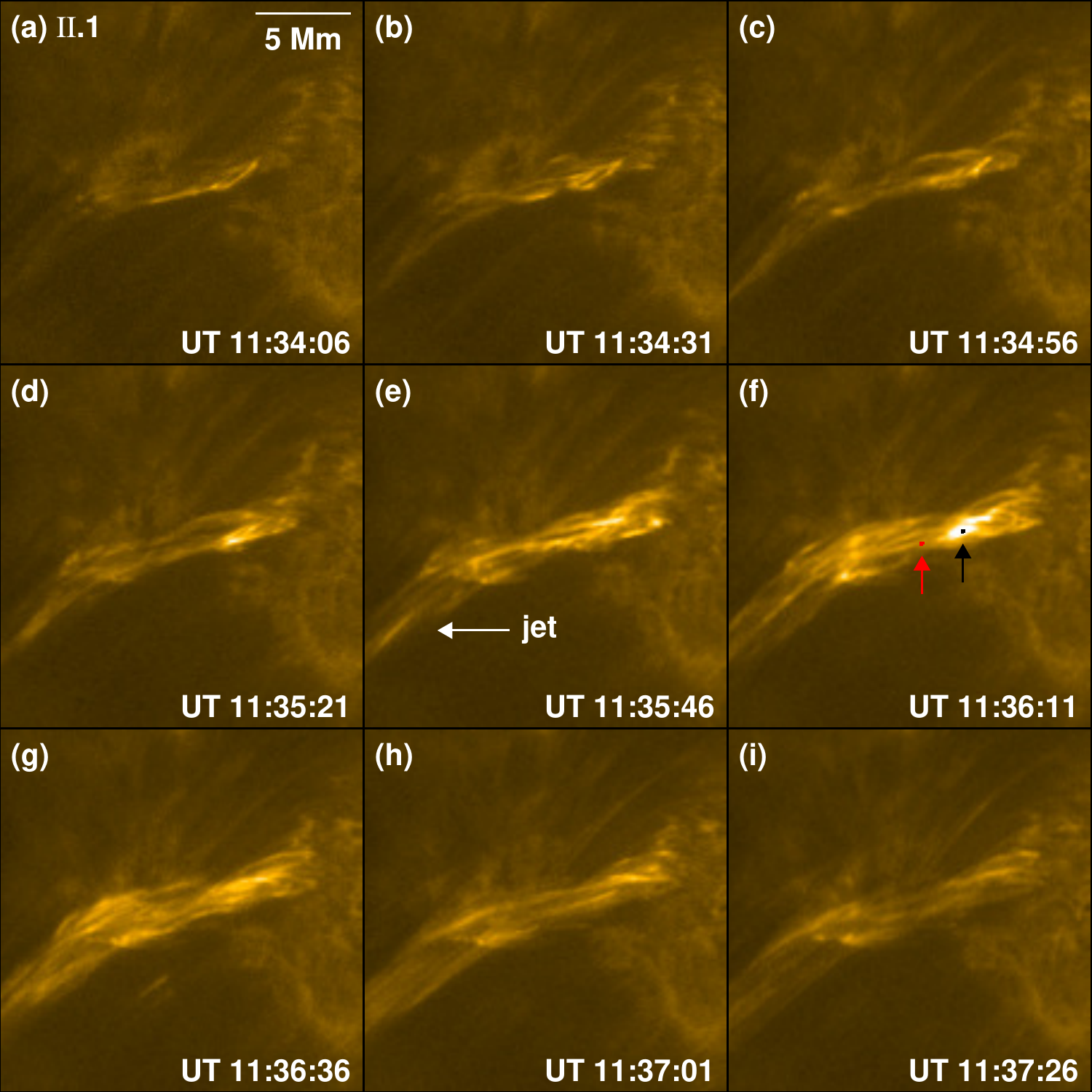}
\includegraphics[width=0.49\textwidth]{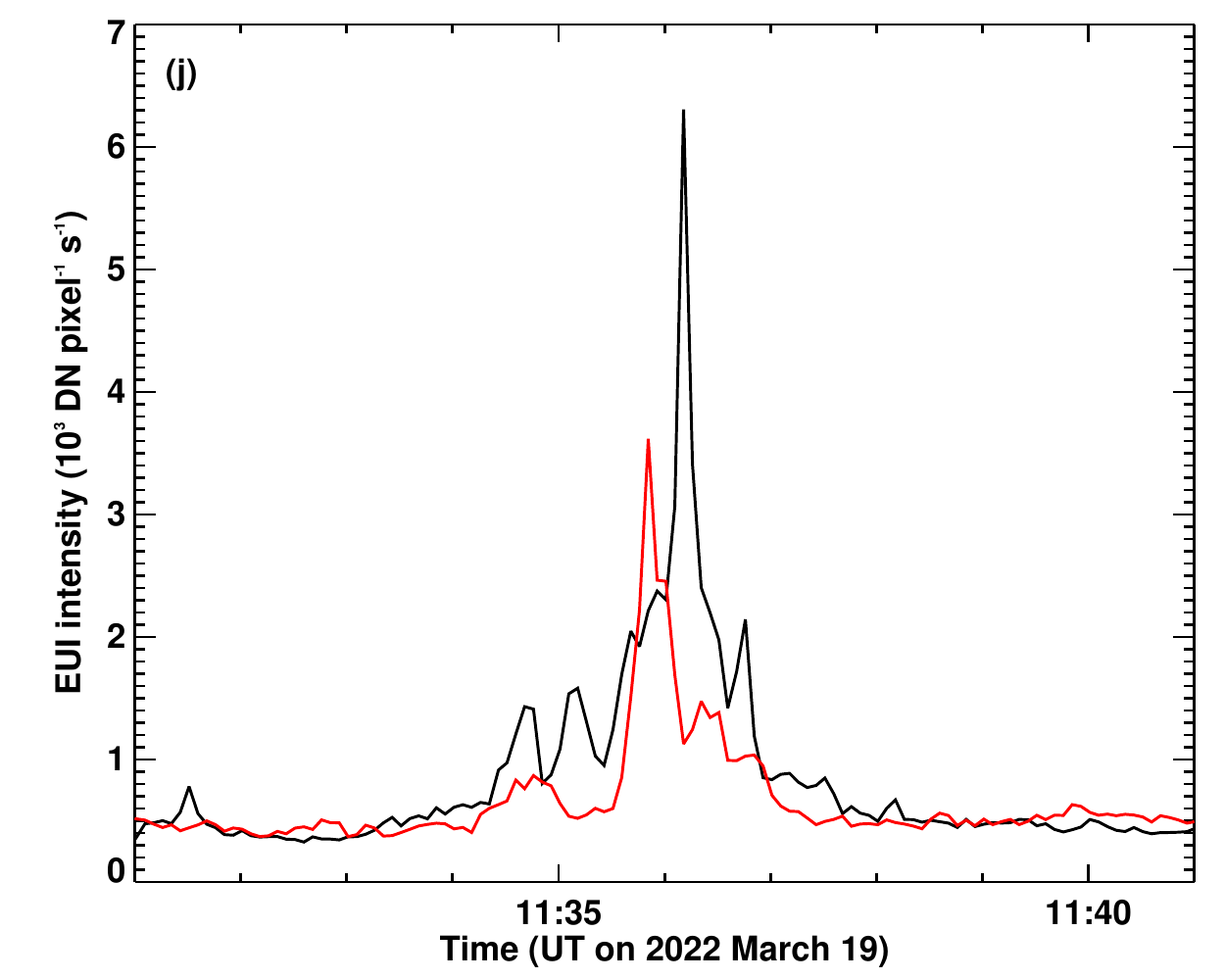}
\caption{Rapidly evolving braids in a compact loop system. Panels (a)--(i) capture the formation and evolution of braided coronal loops in the decaying AR-II (white box in Fig.\,\ref{fig:over2}). In panel (e), the arrow points to a plasma jet. In panel (f), the red and black arrows point to a pair of pixels placed along the braided loops. The red and black \hri\ light curves plotted in panel (j) are obtained from the corresponding colored pixels in panel (f). The intensity of the displayed images is on a square-root scale. See Sect.\,\ref{sec:rapid} for a discussion. An animation of panel a is available online.
\label{fig:braid3}}
\end{center}
\end{figure}

\section{Rapidly evolving braided loops and coronal heating over filamentary structures\label{sec:rapid}}

While the example of braiding presented in Fig.\,\ref{fig:braid1} is associated with conventional coronal loops in the core of an active region, we also identified other braiding-related coronal heating events in AR-I that originate over chromospheric filamentary structures that appear darker in EUV due to absorption. In Fig.\,\ref{fig:braid2} we present such an example of intermittent coronal heating during the untangling of loops over filamentary structures (see box AR-I.2 in Fig.\,\ref{fig:over1}). Unlike the more gentle untangling observed in conventional loops (Fig.\,\ref{fig:braid1}), the loops over filaments presented in Fig.\,\ref{fig:braid2}a untangle more impulsively. Here the SDO/AIA and Hinode/XRT instruments both detected coronal emission (Fig.\,\ref{fig:braid2}b--f) during the untangling of the loops that was clearly seen in \hri\ images (see the animation associated with Fig.\,\ref{fig:braid2}a). Furthermore, nanojet features similar to those discovered by \citet{2021NatAs...5...54A} are also apparent in this case in the \hri\ images. In addition, footpoint brightenings are observed that are similar to the so-called rapid moss variability that is seen in active region cores associated with 5--7\,MK hot loops \citep[e.g.,][]{2014Sci...346B.315T,2018A&A...615L...9C}. The soft X-ray emission as observed by Hinode/XRT is organized in a more loop-like feature, whereas the \hri\ images of the same structure show substantial complexity in both space and time.

This spatial and temporal complexity also extends to the footpoints of these untangling loops (white box in Fig.\,\ref{fig:braid2}a and Fig.\,\ref{fig:braid2}g--j). Here the EUV emission is organized into distinct blob-like features that exhibit back-and-forth motions, including signatures of splitting and merging (red contoured regions in Fig.\,\ref{fig:braid2}g--j). These dynamics could be closely related to the small-angle reconnection scenario of nanojets \citep[][]{2021NatAs...5...54A}. This suggests that the energy deposition at the footpoints is quite bursty and is spatially less coherent than in the braids in conventional loops displayed in Fig.\,\ref{fig:braid1}. 

We observed a similar heating event associated with braided loops over filamentary features on 2022 March 19. In the decaying AR12967 (AR-II; see Fig.\,\ref{fig:over2}), we detected a system of coronal loops that were roughly 10--15\,Mm long and compact. They exhibited rapid evolution, including clear signs of braiding (see Fig.\,\ref{fig:braid3}). In this particular example, we also observed the ejection of confined plasma jets from one end of the loop system. 

As demonstrated in Fig.\,\ref{fig:braid3}a--i, this compact loop system as a whole shows intensity variation over a period of about 180\,s. We observed, however, that individual loops within the system evolve much more rapidly. The system first appears as a single bright loop in the early stages (Fig.\,\ref{fig:braid3}a). Subsequently, many bright loops appear and spread about 1--2\,Mm across the length of the loop. Not only are these loops apparently braided, but the whole system evolves from a nearly untangled state (as seen in EUV images; which does not preclude preexisting tangling in cooler plasma structures in that system) to a highly braided state on short timescales of 60\,s (e.g., compare the development of braided structures from panels a to c to e of Fig.\,\ref{fig:braid3}). This is evident in morphological changes to the structures in subsequent panels that are temporally separated by only 25\,s. The very appearance of apparent tangling in the EUV suggests, however, that energy stored in the preexisting tangling is being dissipated to heat the plasma (see further discussion of the visibility of braids in Sect.\,\ref{sec:disc}).

Similar to Fig.\,\ref{fig:braid2}, this particular braided system also shows more spatial complexity and smaller-scale structuring than the loops presented in Fig.\,\ref{fig:braid1}. In contrast to the nearly simultaneous evolution of \hri\ intensity from regions separated by about 2\,Mm along the loops in Fig.\,\ref{fig:braid1} (see Fig.\,\ref{fig:lc}), two regions from the compact loops with similar separation exhibit distinct intensity variations on shorter timescales (see Figs.\,\ref{fig:braid3}f and \ref{fig:braid3}j)\footnote{Each of the remaining eight pixels surrounding the respective red and black pixels along the braided loops in Fig.\,\ref{fig:braid3} show nearly similar and temporally coherent intensity variations on shorter timescales. This suggests that these intensity variations on shorter timescales that we observed are indeed of solar origin.}. This points to a more spatially and temporally sporadic release of magnetic energy in these compact braided loops. Additional examples of coronal braids over chromospheric filamentary structures are presented in Appendix\,\ref{sec:more}.

\section{Timescales associated with intermittent heating events\label{sec:time}}

The 3\,s\ high-cadence \hri\ data from 2022 March 17 allow us to probe the timescales associated with the thermal response to magnetic reconnection in braiding events. To do this, we compared the case of the braiding within the coronal part of a loop (discussed in Sect.\,\ref{sec:braid}) and the events of the type found over the filamentary structures (Sect.\,\ref{sec:rapid}). 

These two types of events show a clear distinction in their variability.
As shown in Fig.\,\ref{fig:lc}, at the location of localized brightening associated with coronal braids in the conventional loop system, the \hri\ light curves exhibit a smooth rise and fall in intensity over a period of 4--5\,minutes without a clear temporal substructure. The coronal loops over the filaments displayed in Fig.\,\ref{fig:braid2}, however, demonstrate burstyness during untangling, which might be a signature of shorter-term variability in the thermal response of plasma to intermittent heating.

\begin{figure}
\begin{center}
\includegraphics[width=0.49\textwidth]{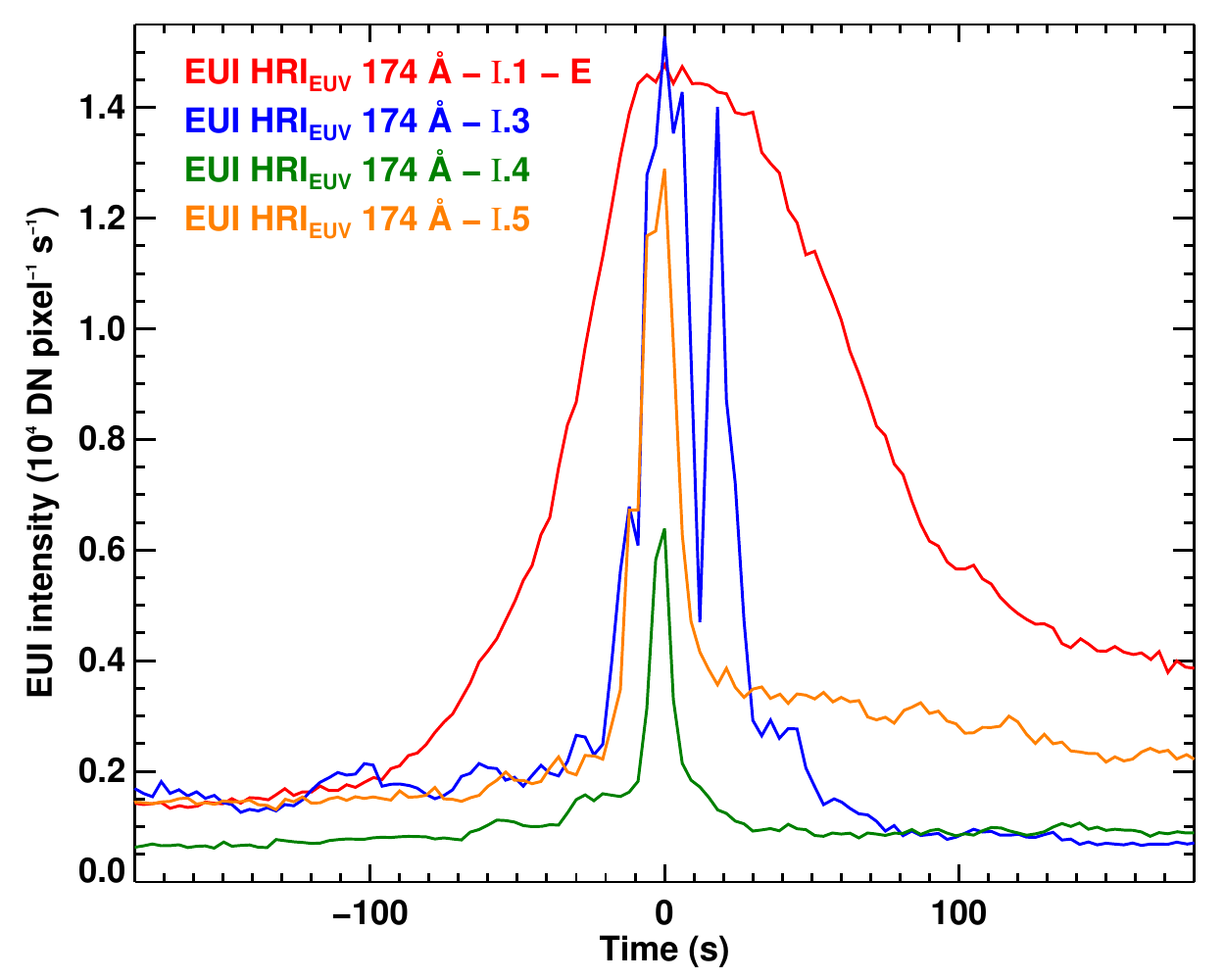}
\caption{Comparison of \hri\ light curves from different heating events. The curve in red is same as that in Fig.\,\ref{fig:lc} (i.e., from point E of the braided loop system in Fig.\,\ref{fig:braid1}), but not normalized to its peak value. The other three light curves (i.e., I.3--I.5) are obtained from the center of the respective panels in Fig.\,\ref{fig:bright}. The 0\,s mark along the abscissa (i.e., the time-axis) is with respect to the peak of each of the light curves. See Sect.\,\ref{sec:time} for a discussion.
\label{fig:lc2}}
\end{center}
\end{figure}

To understand this effect, we selected a few more intermittent heating events from the core of AR-I, which has been observed at the highest cadence. Snapshots of these events are displayed in Fig.\,\ref{fig:bright}, and their spatial relation to the active region core is identified with the respective numbered boxes in Fig.\,\ref{fig:over1}. Events AR-I.3 and AR-I.4 are associated with loops, while event AR-I.5 is a rapid moss variability feature at the footpoints of hotter loops. The light curves from these regions in comparison to the light curve from point E of conventional braided loops in Fig.\,\ref{fig:braid1}a are shown in Fig.\,\ref{fig:lc2}. It is evident that these intermittent heating events AR-I.3--AR-I.5 exhibit temporal variability of emission on much shorter timescales of 10--30\,s than the braided loops in Fig.\,\ref{fig:braid1}a.  This shorter-term variability of EUV emission on timescales shorter than 60\,s is also observed in the compact loop system displayed in Fig.\,\ref{fig:braid3}j. 
Furthermore, considering the footpoint response alone, the moss variability (i.e., the light curve of event AR-I.5 in Fig.\,\ref{fig:lc2}) is more rapid than that observed at the footpoints of the braided loops in Fig.\,\ref{fig:braid1}a (i.e., the black curve in Fig.\,\ref{fig:lc}). These findings suggest that intermittent energy release in the corona through reconnection could have a range of temporal scales from a few 10\,s to a few 100\,s or even longer. The differences in timescales might also arise because the braided loops in Fig.\,\ref{fig:braid1} appear in the aftermath of a B-class flare. The flare itself could correspond to a very different initial magnetic topology, involving many processes,  which in turn might govern the relatively longer timescales of the relaxation of the braid that followed. The data also suggest that the temporal variability might be related to the degree of spatial coherence of energy deposition in the corona, that is, the more temporally intermittent events appear more patchy or bursty in space. However, this would have to be quantified in a future study.

\section{Discussion\label{sec:disc}}
One of the most influential cartoons in solar coronal physics is that of braided loops, as proposed by \citet{1972ApJ...174..499P,1983ApJ...264..642P}. The appearance of the observed compact braided loop system in Fig.\,\ref{fig:braid3} is reminiscent of this cartoon scenario. The question is whether a rapidly evolving braiding feature like this might be used as a test-case to understand how these structures might be generated in the corona. To address this question, we further discuss the evolution of this loop system. 

As described in Sect.\,\ref{sec:rapid}, this particular compact loop system developed tangling on timescales of about 25\,s. We assume that each bright loop within the system is connected to a pair of isolated magnetic field patches at both footpoints in the photosphere. The magnetic field patches in the photosphere move with typical speeds of about 1\,km\,s$^{-1}$ \citep[e.g.,][]{2012ApJ...752...48C}. Then it would take about 1000\,s for a pair of initially parallel loops to exhibit an apparent crisscross pattern within a system that is 1--2\,Mm across. \citet{1988ApJ...330..474P} predicted that for a 10\,Mm-long loop system (comparable to the observed compact loop system in Fig.\,\ref{fig:braid3}), a steady state of energy injection from slow footpoint motions against the release of magnetic energy through reconnection in the tangled coronal magnetic fields is reached in about 5000\,s. The observed braided features in the compact loops (Fig.\,\ref{fig:braid3}) develop and evolve on a much shorter timescale of about 25\,s, however. This indicates that footpoint motions alone would be too slow to directly explain the development of rapid braiding in this system. It is then likely that additional magnetic disturbances, such as the emergence and cancellation of surface magnetic flux, and the associated energy buildup and release through reconnection \citep[][]{2018ApJ...862L..24P,2020A&A...644A.130C}, might be necessary to explain the rapidly evolving system of braided loops. In this context, \citet[][]{2014ApJ...795L..24T} observed that the coronal brightenings in the Hi-C braided loops \citep[][]{2013Natur.493..501C} were externally triggered by photospheric flux cancellation.

Another important point to consider here is that the loops in Fig.\,\ref{fig:braid3} become already visible in the EUV passband that is sensitive to emission from 1\,MK, while the tangling is apparently still developing. This means that the dissipation of energy that causes plasma heating in this system is already in progress. This is possible when part of the preexisting tangling begins to relax through reconnection to heat the plasma. The brightening of loops that are already tangled to begin with might then also give the impression of a rapid development of braiding. In this regard, the coronal braids in all the discussed cases become apparent only for a limited amount of time when the magnetic field is relaxing. This does not, however, preclude the existence or buildup of magnetic tangling in the corona as proposed by \citet{1988ApJ...330..474P}. It only means that the (presumably) slow buildup of magnetic tangling in the corona on timescales of several 1000\,s, in itself, is not observed, at least in the passband we used. The tangling could develop at different temperatures. While the tangling may build up slowly in the corona, the braids become apparent in remote-sensing observations only when a part of the tangled magnetic field is already relaxed to heat the plasma, such that the emission from that heated plasma is captured by a filter (with a given thermal response). However, 3D MHD models predict that an efficient dissipation of currents may prevent magnetic fields from reaching a highly tangled state in the corona \citep[][]{2022ApJ...933..153P}. These 3D MHD models are in general limited by the very high diffusivity, and by design do not allow for the buildup of strong current sheets. Moreover, no spatial and temporal information on the onset of reconnection that relaxes braids in real coronal loops, subjected to slow stressing, is known.

Except for the braided conventional loop system presented in Fig.\,\ref{fig:braid1}, we did not find clear or unambiguous signatures of widespread braiding in longer coronal loops connecting the main opposite-polarity patches of active regions. This might be due to the sensitivity of \hri\ to emission from 1\,MK plasma, implying that the braids could be more apparent in emission from plasma at higher temperatures. As pointed out by \citet{2011ApJ...736....3V}, however, even the X-ray diagnostics that sample emission from higher-temperature plasma do not reveal widespread braiding signatures, but at much lower spatial resolution. Further investigation using numerical simulations is required to determine why only a few loop
systems such as the one presented in Fig.\,\ref{fig:braid1} exhibit observable
braided structures even down to temperatures of 1\,MK while most of the longer
loops do not when the coronal magnetic fields are subjected to continual stressing from slow footpoint motions.

The other coronal braiding examples that we presented here were observed over low-lying chromospheric filamentary structures (see additional discussion in Appendix\,\ref{sec:chrom}). The two braiding examples that \citet{2013Natur.493..501C} discussed also fall into this category. This suggests that chromospheric filaments might play a role in the observability of coronal braids down to temperatures of 1\,MK.

Earlier studies using Hi-C implied that observable signatures of braiding would be common at small spatial scales of about 150\,km \citep[][]{2013Natur.493..501C}. The new EUI observations used here with a spatial resolution of about 250--270\,km, comparable to or better than Hi-C, show that signatures of braiding are not common at these small scales. The reason might either be that the braiding occurs on even smaller scales than previously thought, that it simply eludes observations for the reasons outlined above, or that it is not widespread at all. To this end, the two future missions, the Solar-C EUV High-Throughput Spectroscopic Telescope \citep[][]{2019SPIE11118E..07S} and the Multi-slit Solar Explorer \citep[][]{2022ApJ...926...52D}, could shed light on the multithermal observability of braided coronal loops.

\section{Conclusion\label{sec:conc}}

Using novel high-resolution and high-cadence EUI/\hri\ observations, we presented signatures of intermittent coronal heating during the untangling of braided loops. Our observations suggest that the apparent tangling and subsequent relaxation of braided loops is perhaps more common or observable over chromospheric filamentary structures, although the statistics are so far poor. For one example of a conventional coronal loop system, we showed that the braiding evolved on timescales of minutes and remained observable down to temperatures of 1\,MK. The localized brightening caused by heating at the site of braiding appears to be more coherent spatially (Fig.\,\ref{fig:braid1}). This suggests that a more gentle mode of reconnection might be operating at the location of this coronal braiding. In comparison, more variable and impulsive reconnection might be operating in the braided loops overlying cooler filamentary structure (Figs.\,\ref{fig:braid2}, \ref{fig:braid3}, \ref{fig:braid4}, and \ref{fig:braid5}). In these cases, energy release is likely at smaller spatial scales, which then leads to the formation of blob-like features and rapidly evolving EUV emission from these loops. All these features are also consistent with the interpretation of the recently discovered nanojets \citep[][]{2021NatAs...5...54A, 2022ApJ...934..190S}. While providing evidence for intermittent coronal heating from the relaxation of braided loops, our study identified the operation of both gentle and impulsive modes of energy release through reconnection in the solar corona.

\begin{acknowledgements}
The authors thank the anonymous referee for constructive comments that helped improve the manuscript. L.P.C. thanks Patrick Antolin, Northumbria University, for insightful discussions on nanojets. L.P.C. gratefully acknowledges funding by the European Union. Views and opinions expressed are however those of the author(s) only and do not necessarily reflect those of the European Union or the European Research Council (grant agreement No 101039844). Neither the European Union nor the granting authority can be held responsible for them. S.P. acknowledges the funding by CNES through the MEDOC data and operations center. D.M.L. is grateful to the Science Technology and Facilities Council for the award of an Ernest Rutherford Fellowship (ST/R003246/1). The ROB team thanks the Belgian Federal Science Policy Office (BELSPO) for the provision of financial support in the framework of the PRODEX Programme of the European Space Agency (ESA) under contract numbers 4000134474 and 4000136424. W.L. and M.C.M.C. acknowledge support from NASA's SDO/AIA contract (NNG04EA00C) to LMSAL. Solar Orbiter is a space mission of international collaboration between ESA and NASA, operated by ESA. The EUI instrument was built by CSL, IAS, MPS, MSSL/UCL, PMOD/WRC, ROB, LCF/IO with funding from the Belgian Federal Science Policy Office (BELSPO/PRODEX PEA 4000134088); the Centre National d’Etudes Spatiales (CNES); the UK Space Agency (UKSA); the Bundesministerium f\"{u}r Wirtschaft und Energie (BMWi) through the Deutsches Zentrum f\"{u}r Luft- und Raumfahrt (DLR); and the Swiss Space Office (SSO). AIA is an instrument on board the Solar Dynamics Observatory, a mission for NASA's Living With a Star program. Hinode is a Japanese mission developed and launched by ISAS/JAXA, collaborating with NAOJ as a domestic partner, NASA and STFC (UK) as international partners. Scientific operation of the Hinode mission is conducted by the Hinode science team organized at ISAS/JAXA. This team mainly consists of scientists from institutes in the partner countries. Support for the post-launch operation is provided by JAXA and NAOJ(Japan), STFC (U.K.), NASA, ESA, and NSC (Norway). We are grateful to GOES team for making the data publicly available. This research has made use of NASA’s Astrophysics Data System. 

\end{acknowledgements}

\begin{appendix}
\section{Observational overview\label{sec:app}}

\begin{figure}
\begin{center}
\includegraphics[width=0.49\textwidth]{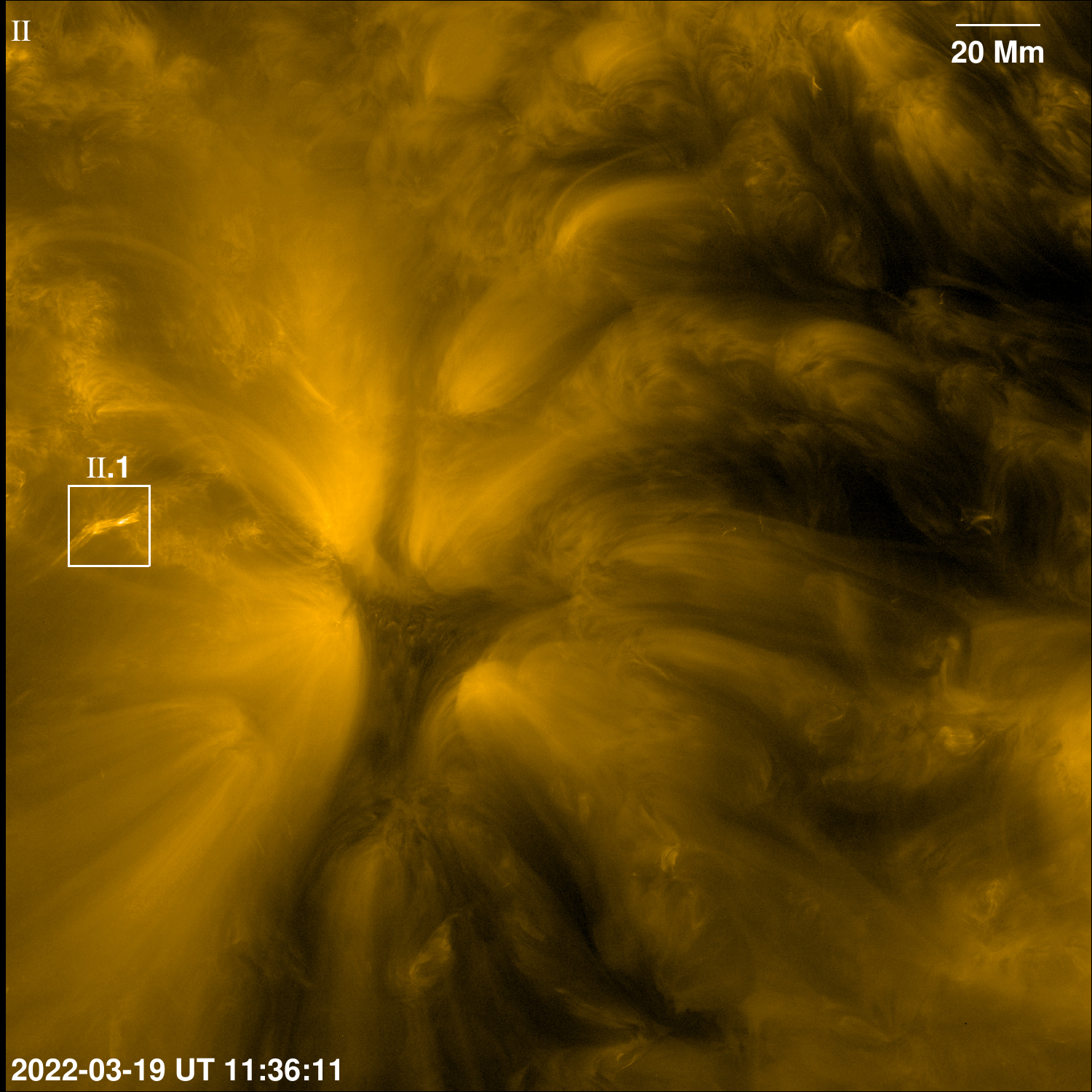}
\caption{Decaying active region AR12967 (AR-II) observed by EUI/\hri\ on 2022 March 19. The white box II.1 outlines a coronal jet event with compact braided loops at its base. The evolution of this loop system is further described in Fig.\,\ref{fig:braid3}. The intensity of the displayed image is on a log scale. North is approximately up. See Table\,\ref{tab:table}, Sect.\,\ref{sec:obs}, and Appendix\,\ref{sec:app} for details.
\label{fig:over2}}
\end{center}
\end{figure}

As mentioned in the main text, we considered three EUI/\hri\ observing campaigns, covering four active regions  (details given in Table\,\ref{tab:table}). At the time of the  \hri\ campaign on 2022 March 17 (Fig.\,\ref{fig:over1}), the evolved active region AR-I was relatively quiescent. Overall, the disk-integrated soft X-ray flux in the 1-8\,\AA\ band of the Geostationary Operational Environmental Satellite (GOES) remained below C-class-flare level. Some 5\,minutes before the start of this \hri\ campaign, however, GOES detected an enhancement of the soft X-ray flux due to a small B-class flare in the active region under investigation (see Fig.\,\ref{fig:goes}). The decaying active region AR-II (Fig.\,\ref{fig:over2}) was quiescent without flaring activity during the time of observations on 2022 March 19. The GOES soft X-ray flux level was therefore lower than that of a C-class flare. The \hri\ captured two active regions (AR12975: AR-III and AR12976: AR-IV) on 2022 April 1 (see Fig.\,\ref{fig:over3}). During that entire day, the Sun was moderately active and produced several C-class flares from many active regions. The GOES soft X-ray background flux was at C level. Some of these flares could be traced to AR-III and AR-IV themselves. During the span of \hri\ observations, however, AR-III was noticeably more active than AR-IV. In summary, our sample of three observing campaigns covers active regions from little to moderate flaring activity. 

\subsection{Aligning \hri\ image sequences\label{sec:align}}
The \hri\ observing sequences include jitter. We removed this jitter by a cross-correlation technique. To this end, we first divided a given \hri\ observing sequence into shorter sequences or overlapping chunks such that the last image of a given shorter sequence was the same as the first image of the following shorter sequence. Using a cross-correlation technique, we then coaligned all images in a given chunk to the first image in that particular shorter sequence because as these chunks overlap, this will result in the alignment of a given observing sequence as a whole to the first image of that sequence.

\begin{figure}
\begin{center}
\includegraphics[width=0.49\textwidth]{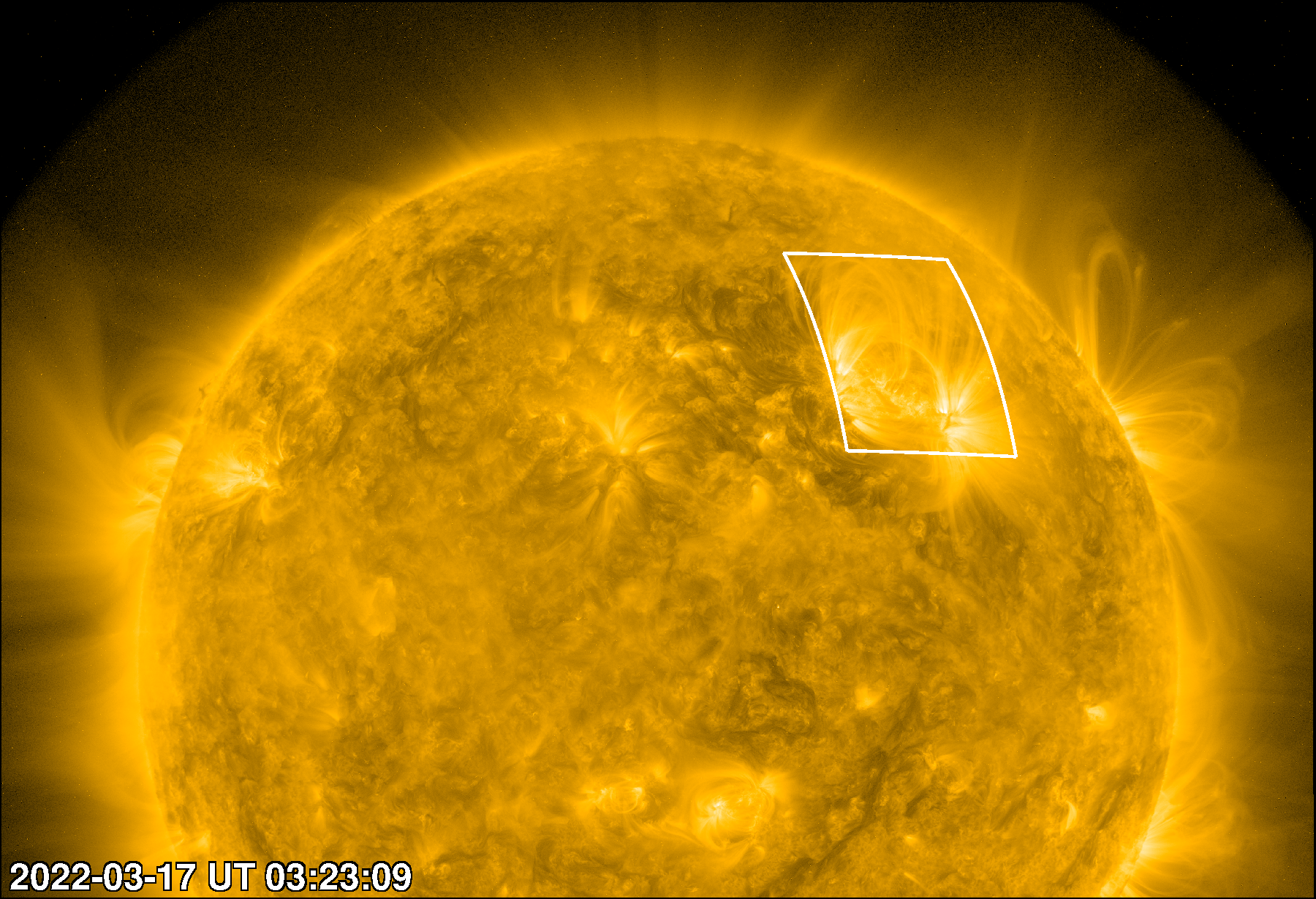}
\caption{Solar corona as observed from the vantage point of Earth with the SDO/AIA 171\,\AA\ EUV filter on 2022 March 17. The white box is the field of view of EUI/\hri, covering AR-I. The displayed snapshot is closest in time with respect to the first image from the time sequence of \hri. During this period, SDO/AIA was operating in a special mode, therefore the roughly lower quarter of the image is not available. The intensity of the displayed image is on a log scale. North is up. See Sect.\,\ref{sec:obs} and Appendix\,\ref{sec:coal} for details.
\label{fig:aia}}
\end{center}
\end{figure}

\begin{figure}
\begin{center}
\includegraphics[width=0.49\textwidth]{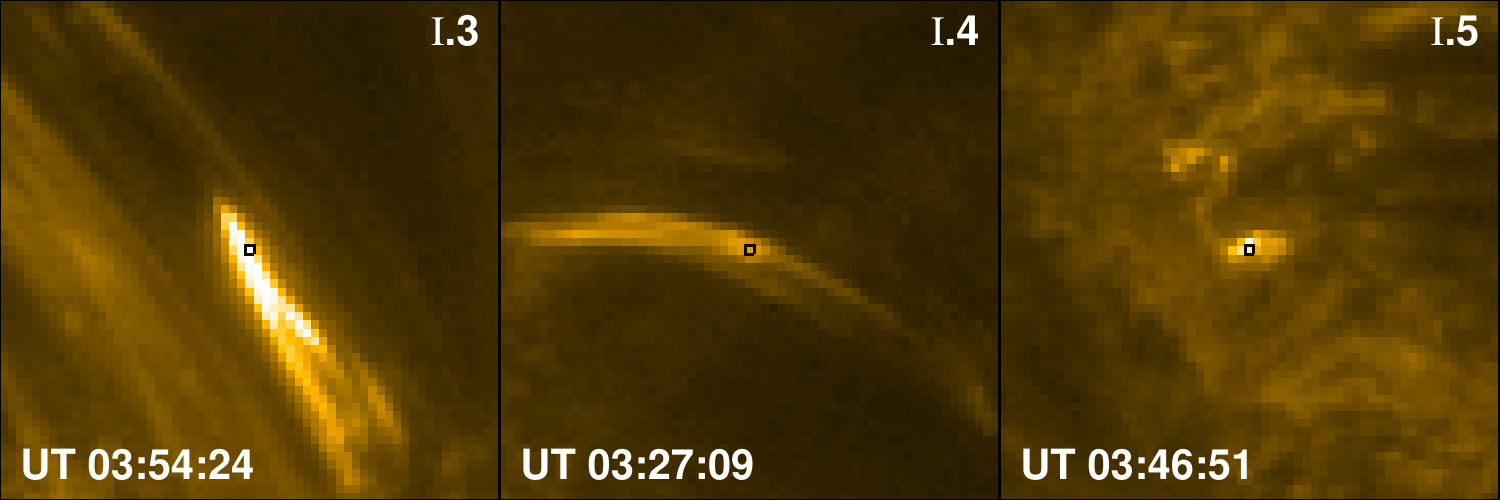}
\caption{Intermittent heating events. Regions I.3--I.5 are outlined in Fig.\,\ref{fig:over1}. The light curves, I.3--I.5, in Fig.\,\ref{fig:lc2} are obtained from the one-pixel-wide black boxes at the center of each panel. The intensity of the displayed images is on a square-root scale. See Sect.\,\ref{sec:time} for a discussion.
\label{fig:bright}}
\end{center}
\end{figure}

\subsection{Co-aligning \hri, AIA, and XRT observations\label{sec:coal}}

During 2022 March 17, the separation angle of Solar Orbiter with Earth was about 27\textdegree. Accordingly, the perspectives of AR-I from the vantage points of Earth and Solar Orbiter were different (see Figs.\,\ref{fig:aia} and \ref{fig:over1}). To retain the high spatial resolution aspect of the \hri\ data, we fixed the first image from the 2022 March 17 sequence as the reference image. To coalign the AIA data to \hri\ images, we first identified the AIA 171\,\AA\ image that was nearest in time to the reference image. Then the AIA data were coregistered and tracked to this nearest-in-time AIA 171\,\AA\ to remove the solar rotation. We then converted the Carrington coordinates of \hri\ reference image into pixel coordinates in the nearest-in-time AIA 171\,\AA\ image. Then the AIA 171\,\AA\ image was re-projected by interpolation onto that pixel coordinates. In principle, this reprojected patch would correspond to the \hri\ field of view as observed from the vantage point of Solar Orbiter. Because of pointing uncertainties of Solar Orbiter, however, there will be some offsets between the \hri\ reference image and the nearest-in-time reprojected AIA 171\,\AA\ image. To obtain these offsets, we cross-correlated the reference and reprojected images to obtain the closest field of view of \hri\ on the AIA 171\,\AA\ image. This offset-corrected field of view of \hri\ is overlaid on the AIA 171\,\AA\ image in Fig.\,\ref{fig:aia}. For the reprojection, we made use of the World Coordinate System (WCS) package available in the solarsoft \citep[see][]{2006A&A...449..791T}. With the offset-corrected pixel coordinates, we reprojected the time sequence of AIA 171\,\AA images. Using this reprojected time sequence, we identified our regions of interest from \hri\ images (Figs.\,\ref{fig:braid1} and \ref{fig:braid2}) in the AIA data.

Because of the separation angle of about 27\textdegree\ of Solar Orbiter with Earth, the AIA data would have to be stretched and distorted by interpolation when reprojected to the \hri\ perspective. This may introduce artifacts in the data. To avoid this, however, the fields of view covering the regions of interest were extracted from the unreprojected AIA data themselves  (i.e., AIA data that are not reprojected to the \hri\ perspective) after
they were identified in the reprojected AIA\ data. This was done by visually inspecting the correspondence of the regions in both the reprojected and unreprojected data. After the regions were identified in the AIA data, it was straightforward to extract corresponding patches from the XRT images as well. 

\subsection{Additional examples of small-scale magnetic braids\label{sec:more}}

\begin{figure}
\begin{center}
\includegraphics[width=0.49\textwidth]{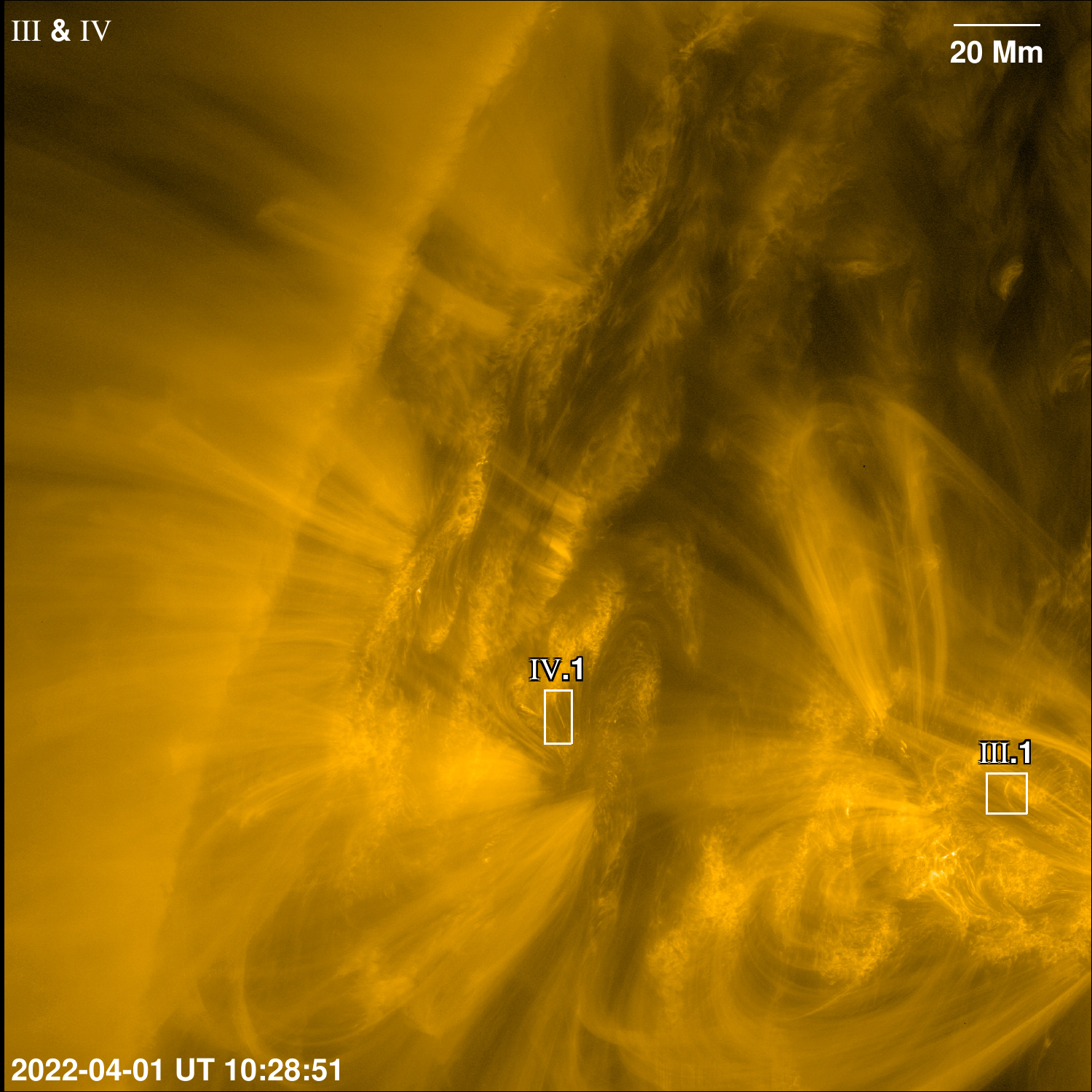}
\caption{Pair of active regions AR-III (western) and AR-IV (eastern) observed by EUI/\hri\ on 2022 April 1. Boxes III.1 (partially covering AR12975) and IV.1 (partially covering AR12976) are further displayed in Figs.\,\ref{fig:braid4} and \ref{fig:braid5}, respectively. The intensity of the displayed images is on a log scale. North is approximately up. See Table\,\ref{tab:table}, Sect.\,\ref{sec:obs}, and Appendix\,\ref{sec:app} for details.
\label{fig:over3}}
\end{center}
\end{figure}

\begin{figure}
\begin{center}
\includegraphics[width=0.49\textwidth]{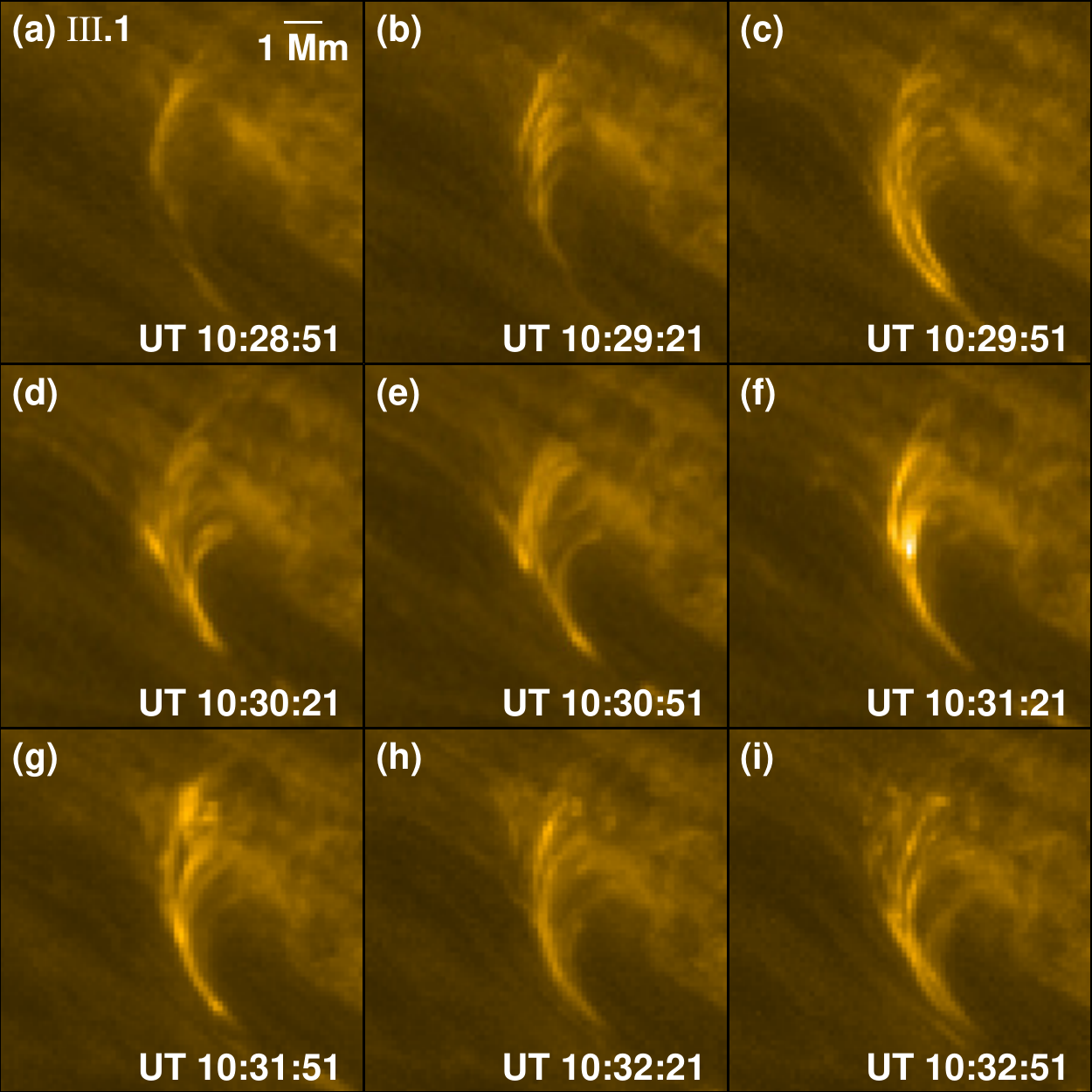}
\caption{Zoomed-in view of region AR-III.1 marked in Fig.\,\ref{fig:over3}. Panels (a)--(i) show the evolution of braided loops over filamentary structures in AR12975. The intensity of the displayed images is on a square-root scale. See Appendix\,\ref{sec:app} for details. An animation of this figure is available online.
\label{fig:braid4}}
\end{center}
\end{figure}

\begin{figure}
\begin{center}
\includegraphics[width=0.49\textwidth]{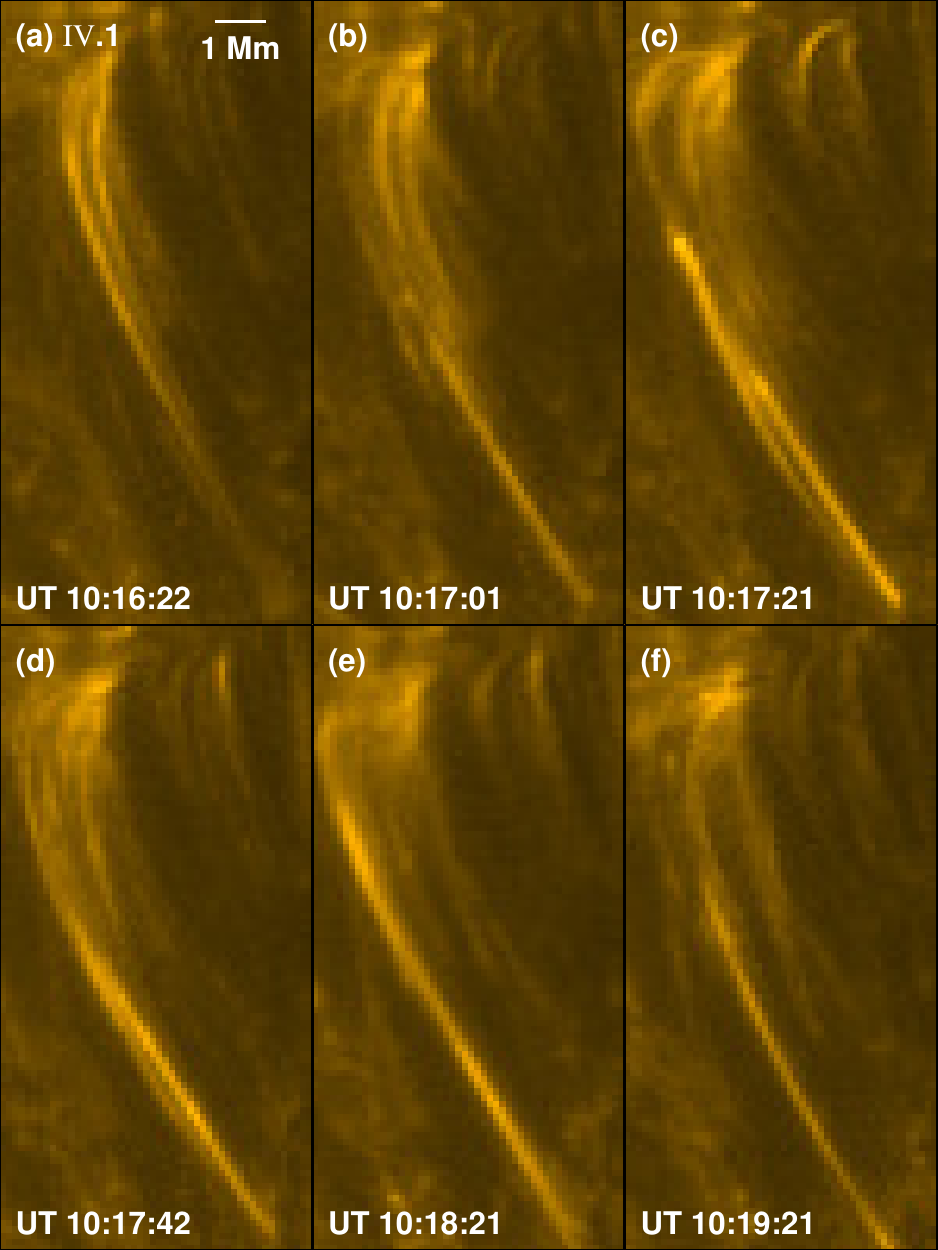}
\caption{Same as Fig.\,\ref{fig:braid4}, but plotted for region AR-IV.1 (AR12976) marked in Fig.\,\ref{fig:over3}. The intensity of the displayed images is on a square-root scale. See Appendix\,\ref{sec:app} for details. An animation of this figure is available online.
\label{fig:braid5}}
\end{center}
\end{figure}

Similar to the braiding examples associated with filamentary structures discussed in the main text (Figs.\,\ref{fig:braid2} and \ref{fig:braid3}), we display in Figs.\,\ref{fig:braid4} and \ref{fig:braid5} additional examples from the \hri\ data of 2022 April 1. The fields of view covered by these regions are marked in Fig.\,\ref{fig:over3}. In Fig.\,\ref{fig:braid4}, we show the rapid development of a braided loop system over a filament in the moderately active AR-III. This loop system shows substantial spatial complexity and evolves on timescales of 30\,s. The morphology of this example closely resembles a braiding case presented in Fig.\,3\ of \citet{2013Natur.493..501C}.

Similarly, the example in Fig.\,\ref{fig:braid5} overlies a penumbral region in the leading sunspot of AR12976. At this location, we observed a number of coronal loops that brightened intermittently. One such event is displayed in Fig.\,\ref{fig:braid5}. A pair of coronal loops are apparently braided, and the morphology of the loop system is shown to evolve on timescales of 20\,s.

\begin{figure*}
\begin{center}
\sidecaption
\includegraphics[width=12cm]{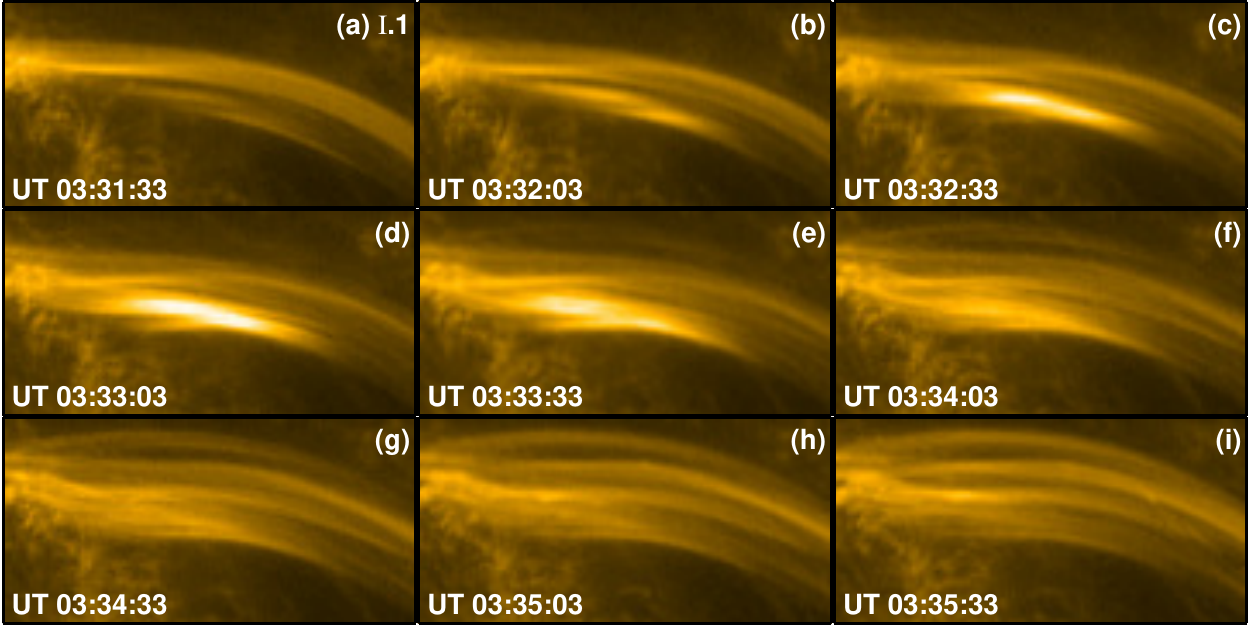}
\caption{Image sequence showing the evolution of braided loops in Fig.\,\ref{fig:braid1}. The subsequent panels are 30\,s apart, and the sequence reveals the relaxation of braided coronal loops from an apparently braided to an untangled state. The intensity of the displayed images is on a square-root scale. See Sect.\,\ref{sec:braid} for a discussion.
\label{fig:seq}}
\end{center}
\end{figure*}

\begin{figure}
\begin{center}
\includegraphics[width=0.49\textwidth]{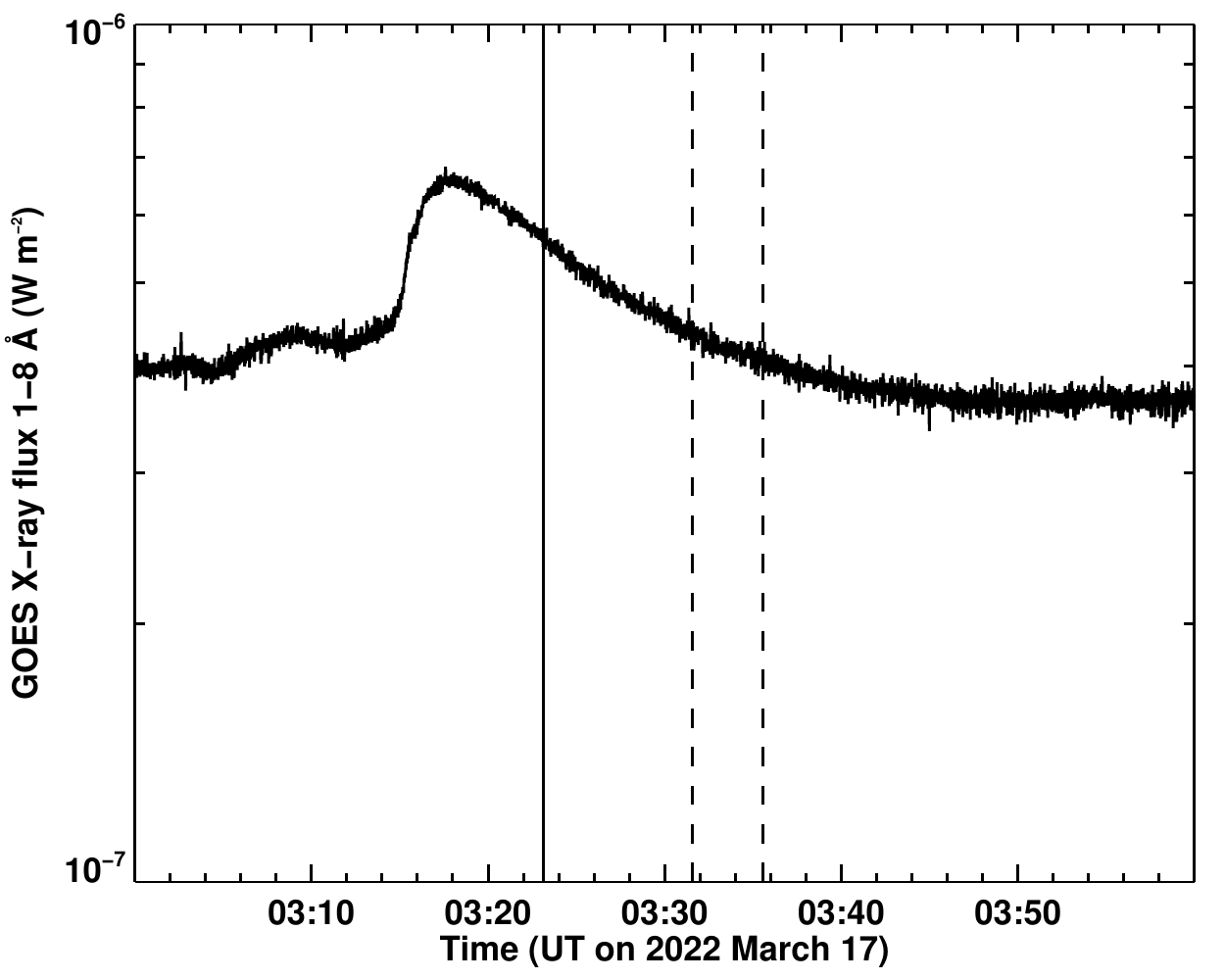}
\caption{Soft X-ray flux. The time series shows the disk-integrated soft X-ray flux detected by the 1-8\,\AA\ band of GOES on 2022 March 17 between UT\,03:00 and UT\,04:00. The vertical solid line marks the start time of EUI observations (corrected for the time delay between Earth and Solar Orbiter; see Table\,\ref{tab:table} for details). The vertical dashed lines mark the duration covered by the panels in Fig.\,\ref{fig:seq}. See Appendix\,\ref{sec:app} for a discussion.
\label{fig:goes}}
\end{center}
\end{figure}

\section{Differential emission measure analysis\label{sec:dem}}

\begin{figure}
\begin{center}
\includegraphics[width=0.49\textwidth]{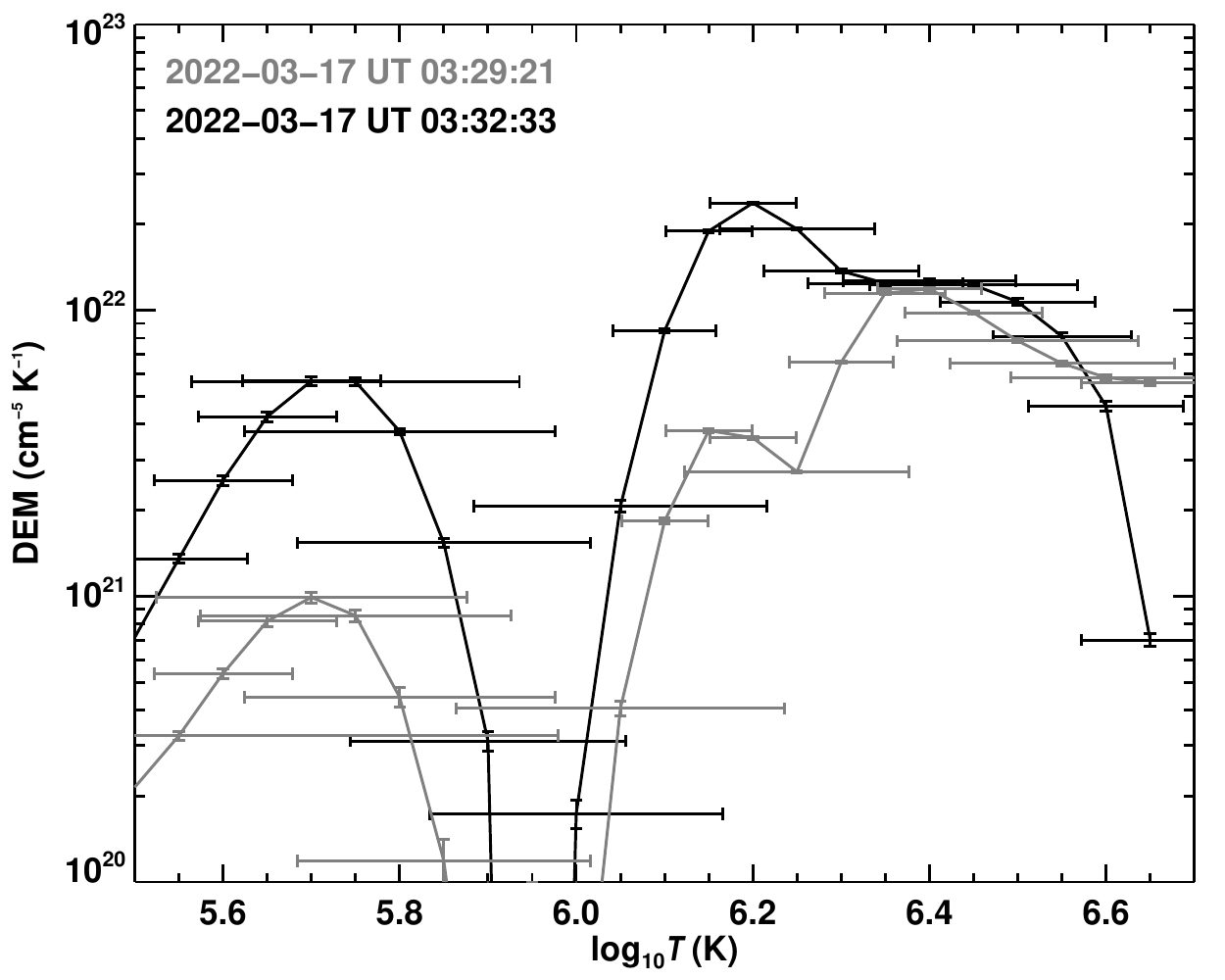}
\caption{Thermal diagnostics of the braided loops. We plot the DEM as a function of temperature for the SDO/AIA mean intensities derived from the braided loops (i.e., the black box on AIA panels in Fig.\,\ref{fig:braid1}). The gray and black curves are computed at time-stamps overlaid on the plot. The vertical error bars correspond to $1\sigma$ errors in the DEM, while the horizontal bars indicate the energy resolution of the regularized inversion technique. See Appendix\,\ref{sec:dem} for a discussion.
\label{fig:dem}}
\end{center}
\end{figure}

We further quantified the thermal evolution of the braided loop described in Fig.\,\ref{fig:braid1} (i.e., AR-I.1), using a differential emission measure (DEM) analysis by employing the aforementioned mean intensities from the six AIA EUV filters (excluding the AIA 304\,\AA\ filter). To this end, we used a regularized inversion technique \citep[][]{2012A&A...539A.146H} to compute the DEM over the temperature range of log$_{10}T$\,(K) 5.5\textendash6.7 at two arbitrary time-steps or phases in the evolution of localized brightening: one at the time close to the peak of the AIA 193\,\AA\ intensity (referred to as the peak phase), and the other at the initial rise phase (these time-stamps are labeled in Fig.\,\ref{fig:dem}). As an input to the inversion scheme, we provided the time-dependent AIA response functions that are calculated by setting the keywords \texttt{chiantifix}, \texttt{evenorm}, and \texttt{noblend} to 1. The AIA EUV uncertainties in intensities were calculated using the \texttt{aia\_bp\_estimate\_error} procedure available in IDL/solarsoft. Furthermore, we set the regularization tweak parameter to 1, and the regularization multiplying factor was set to 1.5. We did not use any initial guess solution. In addition, to aid the method to arrive at a positive solution, the AIA uncertainties were multiplied by a factor 1.2.

The DEMs from these rise and peak phases are plotted as a function of temperature in  Fig.\,\ref{fig:dem}. The DEM at the rise phase (gray curve) shows three local peaks with the global peak around log$_{10}T$\,(K) of 6.4. The DEM curve flattens toward higher temperatures. In comparison, the DEM at the peak phase (black curve) shows the global peak at a lower temperature of log$_{10}T$\,(K) of 6.2. Toward higher temperatures, this DEM shows a plateau of lower emission around log$_{10}T$\,(K) of 6.4 and then a steep fall toward log$_{10}T$\,(K) of 6.6. In addition, the rise phase DEM shows a significantly lower contribution from lower temperatures log$_{10}T$\,(K) of 5.7 in comparison with the peak-phase DEM. The emission measure (i.e., the integral of DEM) is larger for the peak phase than for the rise phase. The overall DEM profiles from these two phases support our interpretation that the localized brightening is associated with multithermal plasma.

\section{Connection to the chromospheric filamentary structures\label{sec:chrom}}

\begin{figure*}
\begin{center}
\includegraphics[width=\textwidth]{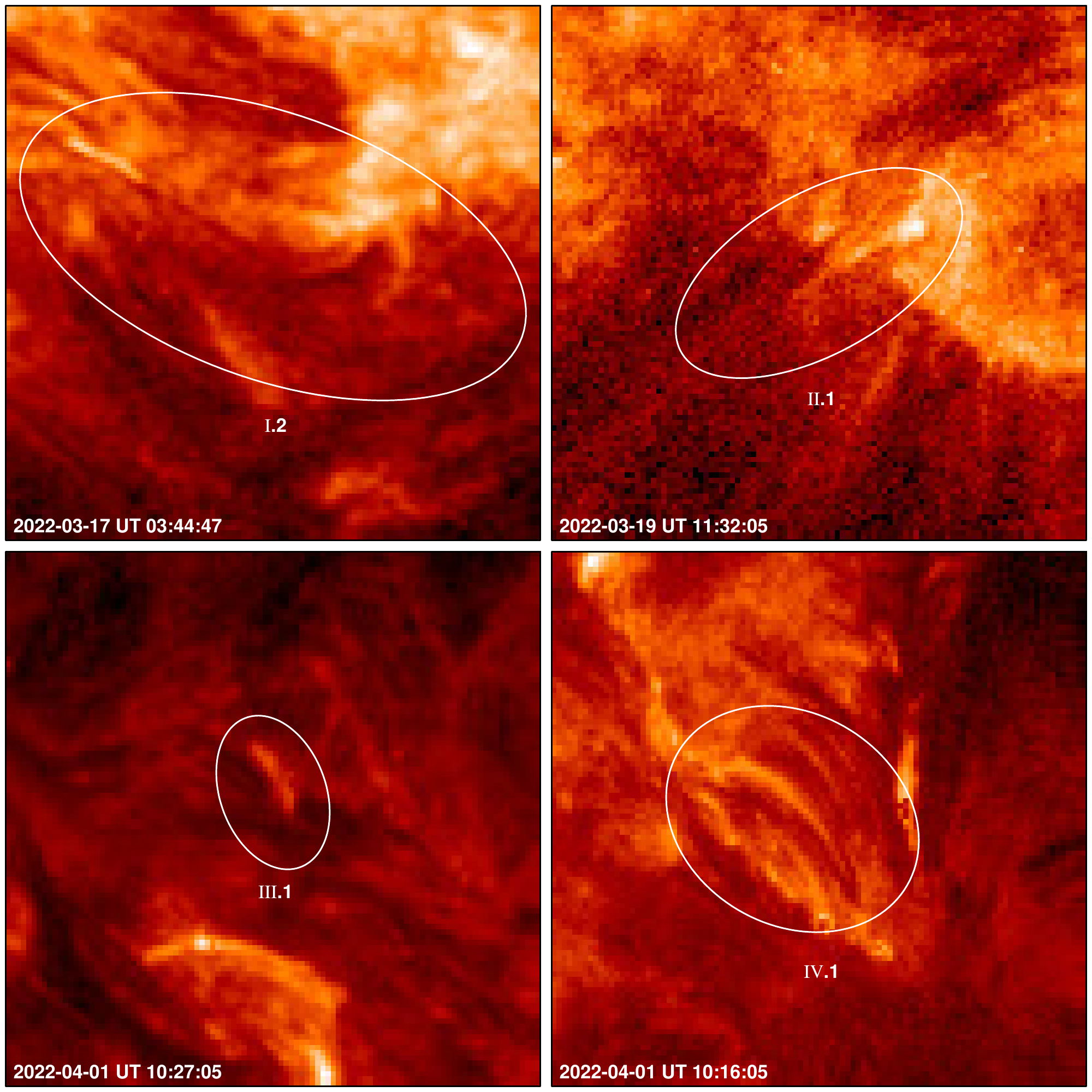}
\caption{Chromospheric filamentary structures.  The four panels show the SDO AIA 304\,\AA\ snapshots of four of the the coronal braids (as labeled in each panel). The ellipses identify the chromospheric filamentary structures associated with the \hri\ braids. Each panel covers a field of view of 44\,Mm\,$\times$\,44\,Mm. North is up. See Appendix\,\ref{sec:chrom} for a discussion.
\label{fig:aia304}}
\end{center}
\end{figure*}

We assess that coronal magnetic braids or at least their observability is more common when they are associated with chromospheric filaments. In our study, four out of five examples clearly overlie cooler filamentary structures (see Figs.\,\ref{fig:braid2}, \ref{fig:braid3}, \ref{fig:braid4}, and \ref{fig:braid5}). To further demonstrate the connection, we made use of the AIA 304\,\AA\ images covering these regions, some minutes before the corresponding \hri\ observations (Fig.\,\ref{fig:aia304}). The regions outlined by ellipses are the locations where we observed signatures of relaxing braided loops in \hri. In all these cases, the AIA 304\,\AA\ images show cooler filamentary material at the locations of \hri\ features. Although not included, the AIA 304\,\AA\ of these features show that the filamentary material exhibits intermittent brightening similar to the \hri\ intensity evolution.

\end{appendix}

\end{document}